\documentclass[numberedappendix]{emulateapj}
\usepackage{amstext}
\usepackage{apjfonts}
\usepackage{amsmath}
\usepackage{amssymb}
\usepackage[colorlinks=true,linkcolor=blue,citecolor=blue]{hyperref}
\usepackage{comment}
\usepackage{mathrsfs}

\usepackage{adjustbox}

\usepackage{enumitem}
\setdescription{font=\normalfont}

\newcommand{\beq}{\begin{equation}}
\newcommand{\eeq}{\end{equation}}

\begin{document}

\title{Topology of Black Hole Binary-Single Interactions} 
\author{Johan Samsing$^{1,2}$, Teva Ilan$^{1}$} 
\altaffiltext{1}{Department of Astrophysical Sciences, Princeton University, Peyton Hall, 4 Ivy Lane, Princeton, NJ 08544, USA}
\altaffiltext{2}{Einstein Fellow}

\begin{abstract} 

We present a study on how the outcomes of co-planar binary-single black hole (BH) interactions distribute as a function
of the orbital initial conditions. We refer to this distribution as the topology. Using an $N$-body code that includes BH finite sizes
and gravitational wave (GW) emission in the equation-of-motion (EOM), we perform more than a million three-body scatterings
to explore the topology of both the classical and the relativistic limit. 
We further describe how the inclusion of GW emission in the EOM naturally leads to scenarios where the binary-single system undergoes
two successive GW mergers. The time span from the first to the second GW merger is shortest in co-planar interactions,
and are for some configurations even short enough for LIGO to observe both events. We note that co-planar interactions could be
frequent in environments such as active galactic nuclei discs.

\end{abstract}

\section{Introduction}

Dynamical interactions of black holes (BHs) in dense stellar environments,
are likely to play an important role for the assembly of binary black hole (BBH) mergers observable by
gravitational wave (GW) interferometers such as LIGO \citep[e.g.][]{2016ApJ...818L..22A, 2016PhRvD..93h4029R, 2016ApJ...831..187A}.
Major effort is especially being devoted to understand the formation and evolution of BHs in
globular cluster (GC) systems \citep[e.g.][]{2014ApJ...784...71S, 2016PhRvD..93h4029R, 2016MNRAS.458.1450W, 2017MNRAS.464L..36A, 2017ApJ...840L..14S, 2017arXiv170603776S}, where
the derived BBH merger rate interestingly seems to be comparable with that inferred from the current sample of
LIGO detections \citep[e.g.][]{2016PhRvL.116f1102A, 2016PhRvL.116x1103A, 2016ApJ...833L...1A, PhysRevLett.118.221101}.
Related few-body studies are therefore often focused on deriving cross sections and rates assuming the orbital and angular momentum
initial conditions (ICs) follow an isotropic distribution at infinity, which would be the case for most dense stellar systems \citep[e.g.][]{Hut:1983js, 2006ApJ...640..156G, 2014ApJ...784...71S,
2016MNRAS.456.4219A, 2016arXiv160909114S, 2017ApJ...840L..14S}.

However, not all few-body interactions take place under isotropic conditions. Dynamical interactions of
BHs in active galactic nuclei (AGN) discs provide one example \citep[e.g.][]{2016ApJ...819L..17B, 2017MNRAS.464..946S, 2017ApJ...835..165B, 2017arXiv170207818M},
where gas torques force objects to move and interact in approximately the same orbital plane as the disc
itself \citep[e.g.][]{1983ApJ...273...99O, 1991MNRAS.250..505S, 2004ApJ...608..108G, 2007MNRAS.374..515L,
2011PhRvD..84b4032K, 2011MNRAS.417L.103M, 2012ApJ...758...51J}. 
Although the number of BHs in AGN discs is currently unknown, it has interestingly been pointed out that if
BHs do exist in such environments, then they are likely to pile up and undergo frequent interactions in migration-traps similar to those found in
planetary discs \citep[e.g.][]{2012ApJ...750...34H, 2012MNRAS.425..460M, 2016ApJ...819L..17B, 2017arXiv170207818M}.
If the BHs have different masses, then the corresponding difference in migration time can also lead to
interactions throughout the disc \citep[e.g.][]{2017arXiv170207818M}.
While this picture is still incomplete and poorly understood, it does provide
at least one example of an environment where few-body interactions do not follow an isotropic distribution.

Motivated by these recent ideas, we explore in this paper the dynamics and outcomes of co-planar binary-single interactions.
For this, we perform a detailed study of the binary-single phase-space, a representation we broadly refer to as the 
binary-single topology in analogy with \cite{1983AJ.....88.1549H}.
Our work should partly be considered as a continuation of that by \cite{1983AJ.....88.1549H}, where the topology of co-planar
binary-single interactions in the Newtonian point-particle limit was studied. In our work we expand on that study by including
General Relativity (GR) corrections in the N-body equation-of-motion (EOM) as well as BH finite sizes.
These corrections break the scale-free nature of the Newtonian point-particle binary-single problem \citep[e.g.][]{1983AJ.....88.1549H, Hut:1983js}, 
leading to several new effects including the formation of GW captures and BH collisions \citep{2014ApJ...784...71S}.
Parts of our presented study describe highly idealized situations, however, we see this as a necessary first step, as the three-body problem
with GR effects is largely unexplored for anisotropic distributions; a proper understanding of the BBH outcome distribution and how it relates to the
binary-single ICs is crucial for connecting observations of GW events to their astrophysical origin.

We further introduce a new unexplored dynamical pathway for generating
two successive GW merger events originating from the same binary-single interaction.
Here the first GW merger happens during the interaction while the three BHs
are still in a bound state \citep[e.g.][]{2014ApJ...784...71S}, where the second GW merger happens between the BH formed in the first merger
and the remaining bound BH. Interestingly, we find the time span from the first to the second
GW merger in co-planar interactions to occasionally be short enough for LIGO to observe both GW mergers.
This suggests that AGN discs might be a likely birthplace for such double GW merger events.

The paper is organized as follows. In Section \ref{sec:Black Hole Binary-Single Interactions}
we introduce the binary-single interaction channel, together with the ICs and outcomes
that are relevant for this study. After this follows in Section \ref{sec:Binary-Single Topology in the Hard Binary Limit} a detailed study of
the binary-single topology for both the classical and the relativistic limit.  
In Section \ref{sec:The Role of GW radiation} we explore in greater detail how the inclusion of GW emission in the N-body EOM
affects the resultant topology and range of outcomes. Results from zoom-in simulations performed to study micro-topological
structures are given in Section \ref{sec:Topological Microstructures}.
Finally, in Section \ref{sec:Double GW Inspirals} we present our proposed channel for generating double GW merger events
through binary-single interactions. Conclusions are given in Section \ref{sec:Conclusions}

\section{Black Hole Binary-Single Interactions}\label{sec:Black Hole Binary-Single Interactions}

In this work we study interactions between a circular binary and an incoming single,
and how the corresponding outcomes distribute as a function of the ICs.
Throughout the paper we assume the three objects to be identical BHs each with mass $m = 10M_{\odot}$,
and corresponding Schwarzschild radius $R_{\rm s}$. All interactions are performed with our $N$-body code
presented in \cite{2016arXiv160909114S}, which includes GR corrections using the
post-Newtonian (PN) formalism \citep[e.g.][]{2014LRR....17....2B}. For the results presented in this paper, we include the 2.5PN
term for describing the dynamical loss of energy and angular momentum through GW emission. The lower order energy preserving precession
terms are not included for simplicity.

In this section we first summarize a few dynamical properties of binary-single interactions
with GW emission and finite sizes included in the EOM.
We then list the ICs relevant for our considered binary-single study together with the range of possible outcomes.
The last part contains a description of what we refer to as the topology. We refer the reader to \cite{1983AJ.....88.1549H} for
further details.

\subsection{Three-Body Energetics and Interactions}\label{sec:Hard-Binary Interactions}

The range of possible outcomes is broadly determined by
the orbital velocity of the initial binary, $v_{\rm orb}$, relative to the velocity of the incoming single
at infinity, $v_{\infty}$. In the equal mass case, one finds
that the initial orbital binary-single energy is negative for $v < 1$ and positive for $v > 1$, where
$v \equiv v_{\infty}/(v_{\rm orb}\sqrt{3})$, as further described in \cite{Hut:1983js}.
These two limits are usually referred to as the hard binary (HB) limit and the soft binary (SB) limit, respectively \citep[e.g.][]{Heggie:1975uy}.
We describe a few characteristics of these two limits below.
We will argue that only the HB limit is relevant for our study. 

\subsubsection{Soft-Binary Limit}

In the SB limit the binary-single interaction is always prompt \citep{1983ApJ...268..342H},
any dynamical corrections to the $N$-body EOM will therefore rarely have a chance to affect
the standard Newtonian distribution of outcomes. As a result, the inclusion of GW emission, as well as collisions
and tidal interactions, does not play a significant role in this limit \citep[e.g.][]{2014ApJ...784...71S}.
The topology of the SB limit was studied by \cite{1983AJ.....88.1549H}, and will therefore not be
discussed further in this paper.

\subsubsection{Hard-Binary Limit}\label{sec:Hard-Binary Limit}

In the HB limit the binary-single system either undergoes a direct interaction (DI) or
a resonant interaction (RI), depending on the exact ICs \citep[e.g.][]{2014ApJ...784...71S}.
For equal mass interactions, these two interaction types appear statistically in near equal numbers.
However, the RI channel has a much higher probability than the DI channel of entering a state where GR effects
and finite sizes can affect the three-body dynamics, as described in the following.

A DI is characterized by a duration time that is similar to the orbital time of the initial binary,
which makes it somewhat similar to a SB limit interaction. The kinematics of a DI is often such that the incoming
single promptly pairs up with one of the binary members through a sling-shot interaction that ejects the remaining member.
The short nature of this interaction type implies that GR effects and collisions rarely have a chance to affect the outcome.

A RI lasts instead from a few to several thousand of initial orbital times \citep[e.g.][]{Hut:1983js}.
The inclusion of GR effects and finite sizes in the EOM therefore have a relative high chance of affecting the dynamics
during this type of interaction. How this in particular leads to an increase in the number of BH mergers can be understood as follows. 
As envisioned by \cite{2014ApJ...784...71S}, a RI can be described as a series of
intermediate states (IMS), each of which is characterized by an IMS binary with a bound single.
Due to pair-wise semi-chaotic exchanges of energy and angular momentum during the resonating evolution, each
IMS binary has a finite probability for being formed with a high eccentricity, even when the initial binary is circular. If the eccentricity is
high enough, GR will affect the EOM of the IMS binary \citep[e.g.][]{Peters:1964bc}, which most notably
leads to an inspiral driven by GW emission that is followed by a prompt merger \citep{2006ApJ...640..156G, 2014ApJ...784...71S}.
For even higher eccentricities the IMS binary members will instead undergo a prompt collision \citep[e.g.][]{2016arXiv160909114S, 2017arXiv170603776S}.

\subsection{Initial Conditions}\label{sec:Initial Conditions}

\begin{figure}
\centering
\includegraphics[width=\columnwidth]{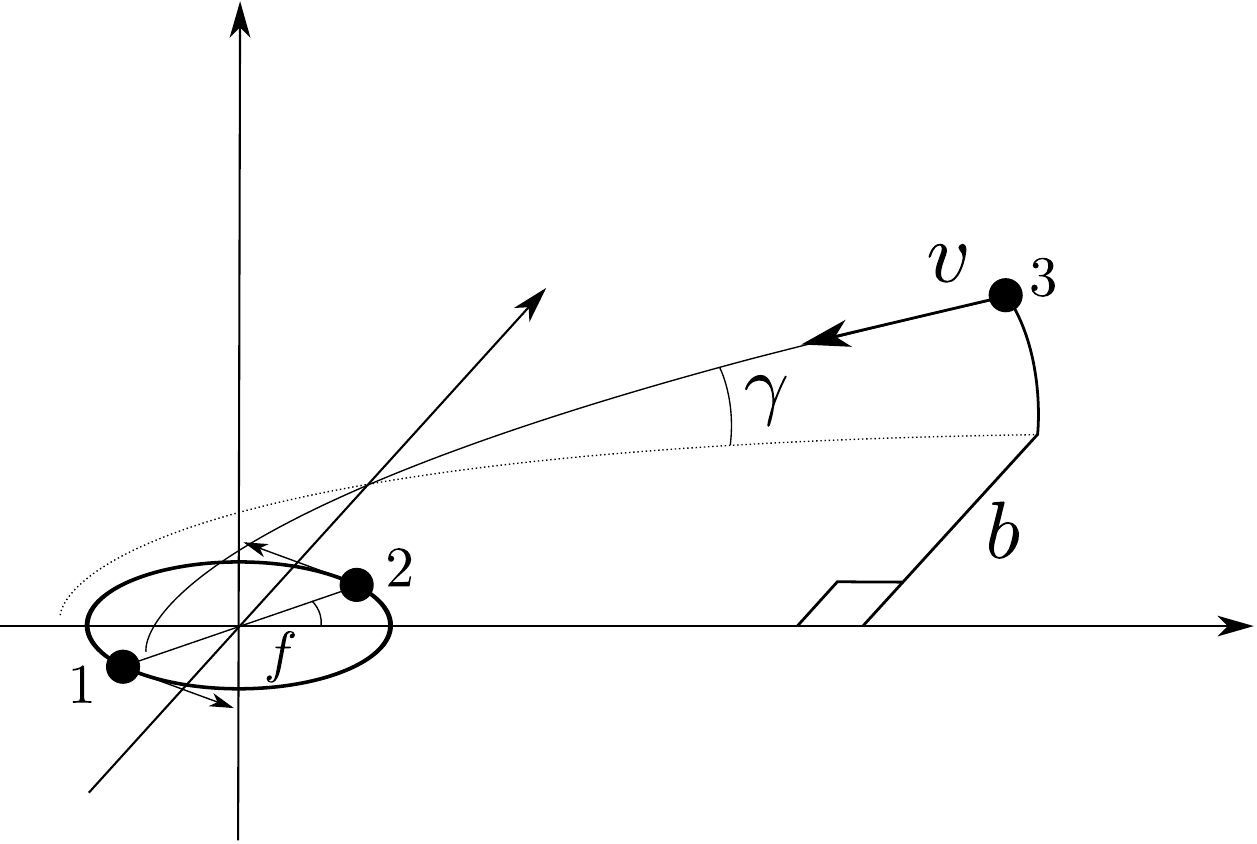}
\caption{Initial conditions and notations for binary-single scatterings. The black dots represent the three interacting BHs, where the indices `1' and `2' denote the initial target
binary members, and `3' the incoming single. The impact parameter $b$ and the relative velocity $v$, are both defined at infinity.
The angle $f$ denotes the orbital phase of the binary at the time the distance between the binary COM and the incoming single is exactly $20 \times a_{0}$, where
$a_{0}$ is the initial binary SMA. The angle $\gamma$ denotes the angle of the incoming single at infinity relative to the initial
binary orbital plane. In this work we study how the outcome of a binary-single interaction depends on $f,\ b$ for fixed values of $v,\ \gamma$.
The initial conditions are further described in Section \ref{sec:Initial Conditions}.
}
\label{fig:scattering_ill}
\end{figure}

The IC parameters that are relevant for our presented binary-single study, are listed and
described in the following. An accompanying illustration is shown in Figure \ref{fig:scattering_ill}.

\begin{description}

\item[\textnormal{Semi-major axis}]
($a_{0}$). The semi-major axis of the initial target binary is denoted by $a_{0}$. As described later,
both the classical and the relativistic limit can be studied by simply varying the value of $a_{0}$.

\item[\textnormal{Velocity}]
($v$). We denote the rescaled relative velocity between the initial
binary COM and the incoming single at infinity by $v$, as described in Section \ref{sec:Hard-Binary Interactions}.
In this work we set $v=0.01$ for all our scatterings, however, none of our conclusions
depend on the exact value of $v$, as long as $v\ll1$.

\item[Binary phase]
($f$). The orbital phase of the binary when the single is exactly $20 \times a_{0}$ away from the binary COM
is denoted by $f$. The value of $f$ is defined relative to an axis that is parallel
to the velocity vector of the incoming single at infinity.

\item[Impact parameter]
($b$). We define the rescaled impact parameter at infinity by $b \equiv b_{\infty}/b_{c}$,
where $b_{\infty}$ is the physical value and $b_{c} = a_{0}/v$. The value of $b$ can be positive or negative,
where a positive (negative) $b$ here corresponds to that the angular momentum vectors of the binary and the incoming single, respectively, have the same (opposite) sign. 
The initial three-body angular momentum in the co-planar case will therefore equal zero for $b = -\sqrt{3/4} \approx - 0.87$ \citep[e.g.][]{1983AJ.....88.1549H}.
We further note that the pericenter distance of the single w.r.t. the binary COM, $r_{\rm p}$, relates to $b$ as $r_{\rm p}/a_{0} = 3b^{2}/8$. 

\item[Orbital plane angle]
($\gamma$). The angle between the orbital planes of the initial binary and the incoming single
is denoted by $\gamma$. In this notation, a co-planar interaction will have $\gamma = 0$.

\end{description}

\subsection{Outcomes and Kinematical Properties}\label{sec:Definition and Identification of Endstates}

The set of binary-single outcomes and kinematical properties we keep track of in this work, is listed and described below. Further details on defining
and identifying outcomes in $N$-body simulations can be found in, e.g., \cite{Fregeau:2004fj, 2014ApJ...784...71S, 2016arXiv160909114S}.

\subsubsection{Outcomes}\label{sec:Definition and Notation of Endstates}

\begin{description}

\item[Binary-single]
(BS[ij]). The most common endstate is a binary with an unbound single. We denote this endstate by BS[ij],
where `BS' is short for `Binary-Single', and [ij] refers to the object pair in the final binary.
Throughout the paper we denote the two initial binary members by `1' and `2', and the incoming single by `3'.

\item[GW inspiral]
(GW). A close passage between any two of the three interacting objects,
occasionally leads to a GW capture, where the two binary members undergo an inspiral
through the emission of GWs \citep[e.g.][]{2006ApJ...640..156G, 2014ApJ...784...71S}.
We generally refer to this endstate by a `GW inspiral', but will in our figures simply use `GW' to keep the
notation short.

\item[Collision]
(Coll). If two objects pass each other at a distance smaller
than the sum of their radii without undergoing a GW inspiral first, then the outcome
is denoted a `collision', or in short by `Coll'. In our case, the collisional distance is simply $2R_{\rm s}$.

\item[Very long interaction]
(VLint). A finite fraction of the interactions can in theory last infinitely long time. This happens generally when the single in an IMS
is sent out on a nearly unbound orbit.
In this work we denote an interaction that did not finish within $2500 \times T_{\rm orb}$ by `VLint' , where $T_{\rm orb}$ is the orbital time
of the initial target binary --  with the only exception of the results shown in Figure \ref{fig:topmap_1exam}, where the limit
is set to $1000 \times T_{\rm orb}$.

\end{description}

\subsubsection{Kinematical Properties}

\begin{description}

\item[Number of intermediate states]
($N_{\rm IMS}$). For each interaction we count the number of times the binary-single
system splits into an IMS. The total number from initial interaction to final outcome is what we denote $N_{\rm IMS}$.

\item[Minimum distance]
($R_{\rm min}$). For all interactions ending as a BS[ij] endstate, we save the minimum distance
any two of the three objects undergo during the interaction. This distance is denoted by $R_{\rm min}$.

\item[Binary-single energy]
($E_{\rm BS}$). At the end of each interaction, we calculate the orbital energy
of the single and the COM of the remaining pair. We denote this energy by $E_{\rm BS}$. In this notation, a BS[ij] endstate will have $E_{\rm BS} > 0$, where
collisions and GW inspirals generally have $E_{\rm BS} < 0$, as they preferentially form during the interaction.

\end{description}

\subsection{Topological Mappings}\label{sec:Topological Mappings}

\begin{figure*}
\includegraphics[width=\textwidth]{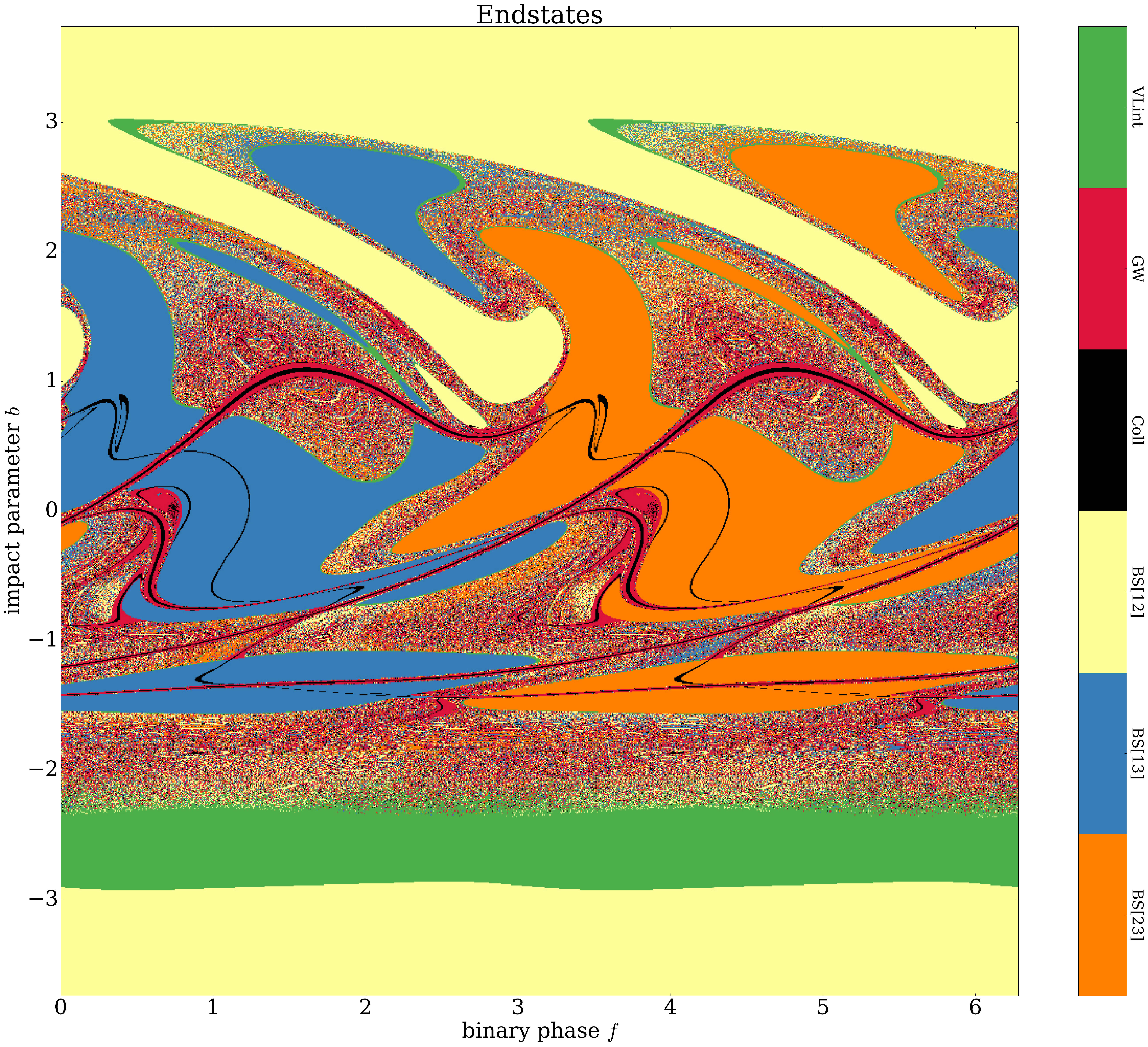}
\caption{Topological map of binary-single endstates. The figure shows the distribution of endstates derived from equal mass BH binary-single
interactions, as a function of the binary phase $f$ and impact parameter $b$, for BH mass $m=10M_{\odot}$, initial SMA $a_{0}=10^{-2}$ AU, relative velocity $v = 0.01$, and 
orbital plane angle $\gamma = 0.0$. We generally refer to this distribution as the binary-single endstate topology, as described in Section \ref{sec:Topological Mappings}.
The large-scale patterns and structural regions seen in the figure, clearly illustrate that the outcome of a binary-single interaction is not completely random,
although the three-body problem is known to be chaotic. In this work we use topological mappings, similar to the one shown here, as a tool to gain further insight
into the binary-single problem with finite size effects and GR corrections included in the three-body EOM.
}
\label{fig:topmap_1exam}
\end{figure*}

The graphical representation of the distribution of binary-single outcomes as a function of the
ICs, is what we loosely refer to as the topology, in analogy with \cite{1983AJ.....88.1549H}.
As an example, Figure \ref{fig:topmap_1exam} shows the distribution of binary-single endstates
as a function of $f,\ b$, for $a_{0}=10^{-2}$ AU and $\gamma = 0$. In our terminology,
we refer to this distribution as the endstate topology, or the endstate-map.
Building on the work of \cite{1983AJ.....88.1549H}, we focus in this study on topological maps derived from
varying $f$ and $b$, while keeping $v$ and $\gamma$ fixed.

\section{Topology of Binary-Single Interactions}\label{sec:Binary-Single Topology in the Hard Binary Limit}

In this section we explore the binary-single topology on large-scales.
For this, we first study the {\it classical limit}, which represents the idealized case where the
objects are all point-like particles interacting through Newtonian gravity only.
After this follows a study of the {\it relativistic limit}, in which GR corrections and finite size effects
start to affect the classical topology.
In the last part we briefly explore how the topological maps change for varying $\gamma$.

For all scatterings performed in this section we keep finite sizes and GW emission in the EOM for consistency;
the classical and the relativistic limit are therefore studied by simply varying $a_{0}$.
We note that $f$ conveniently happens to approximately be
the phase of the binary as the incoming single enters the binary. For example, at $f \approx 0,\ b \approx 0$,
we find that all three objects are approximately on a line as the single enters. In our following descriptions we will therefore
simply refer to $f$ instead of the exact binary phase as the single enters the binary. Furthermore, we often refer to just a single value of $f$,
however, due to the rotational symmetry of the binary, results are similar for any phase $f + n\pi$, where $n$ is an integer.

\subsection{Classical Limit}\label{sec:Classical Limit}

\begin{figure*}
\centering
{\includegraphics[width=\columnwidth]{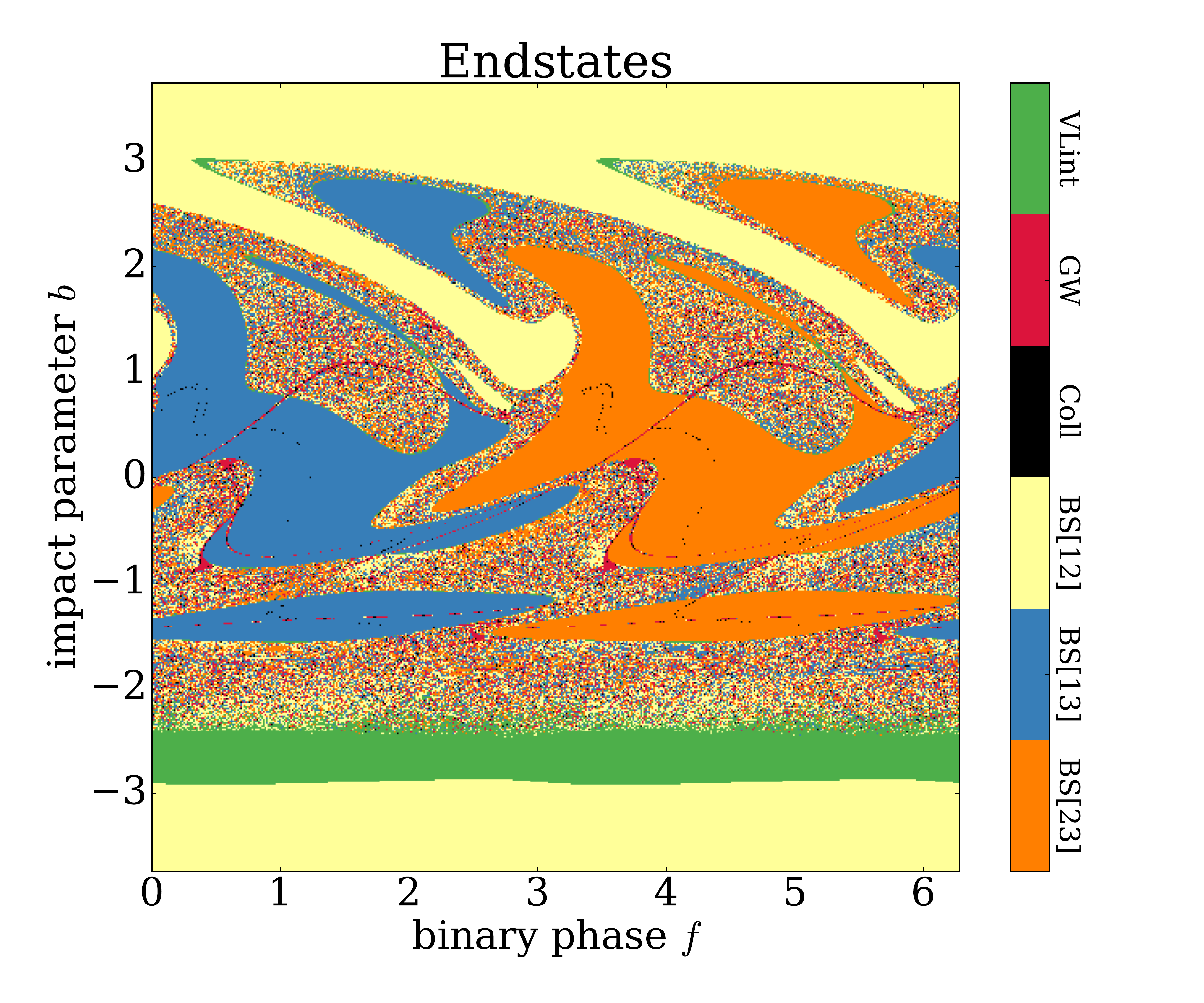}}
{\includegraphics[width=\columnwidth]{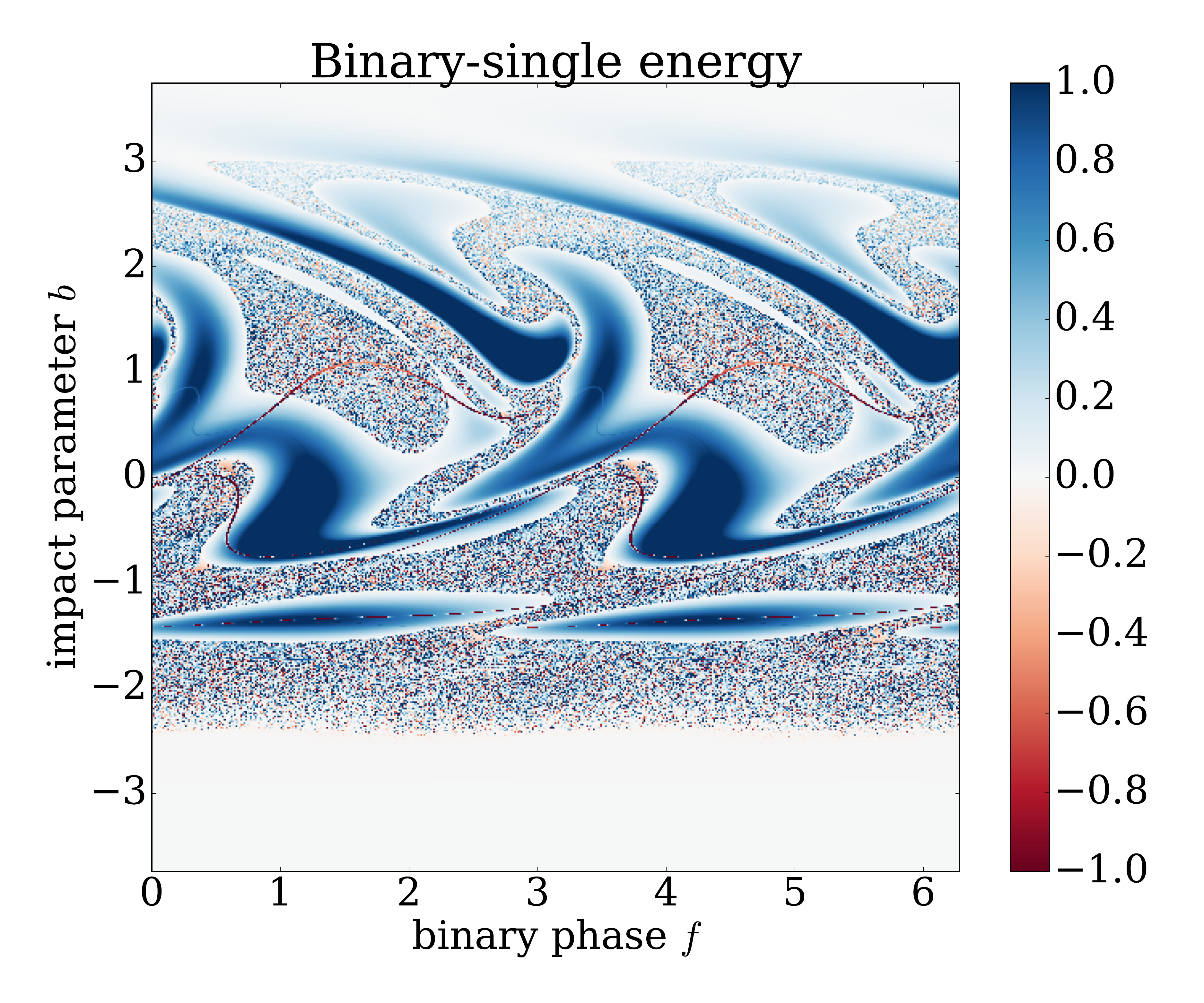}}
{\includegraphics[width=\columnwidth]{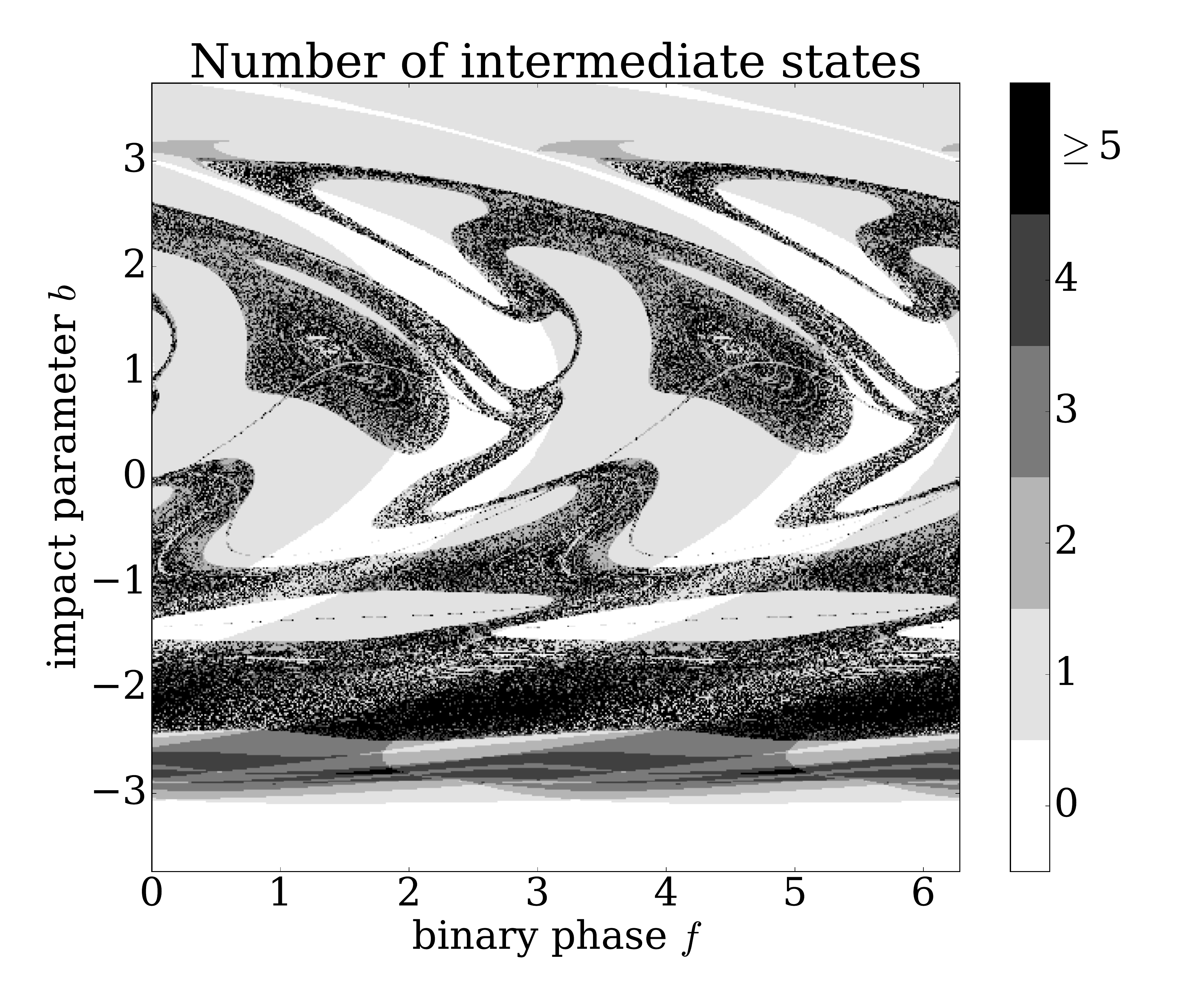}}
{\includegraphics[width=\columnwidth]{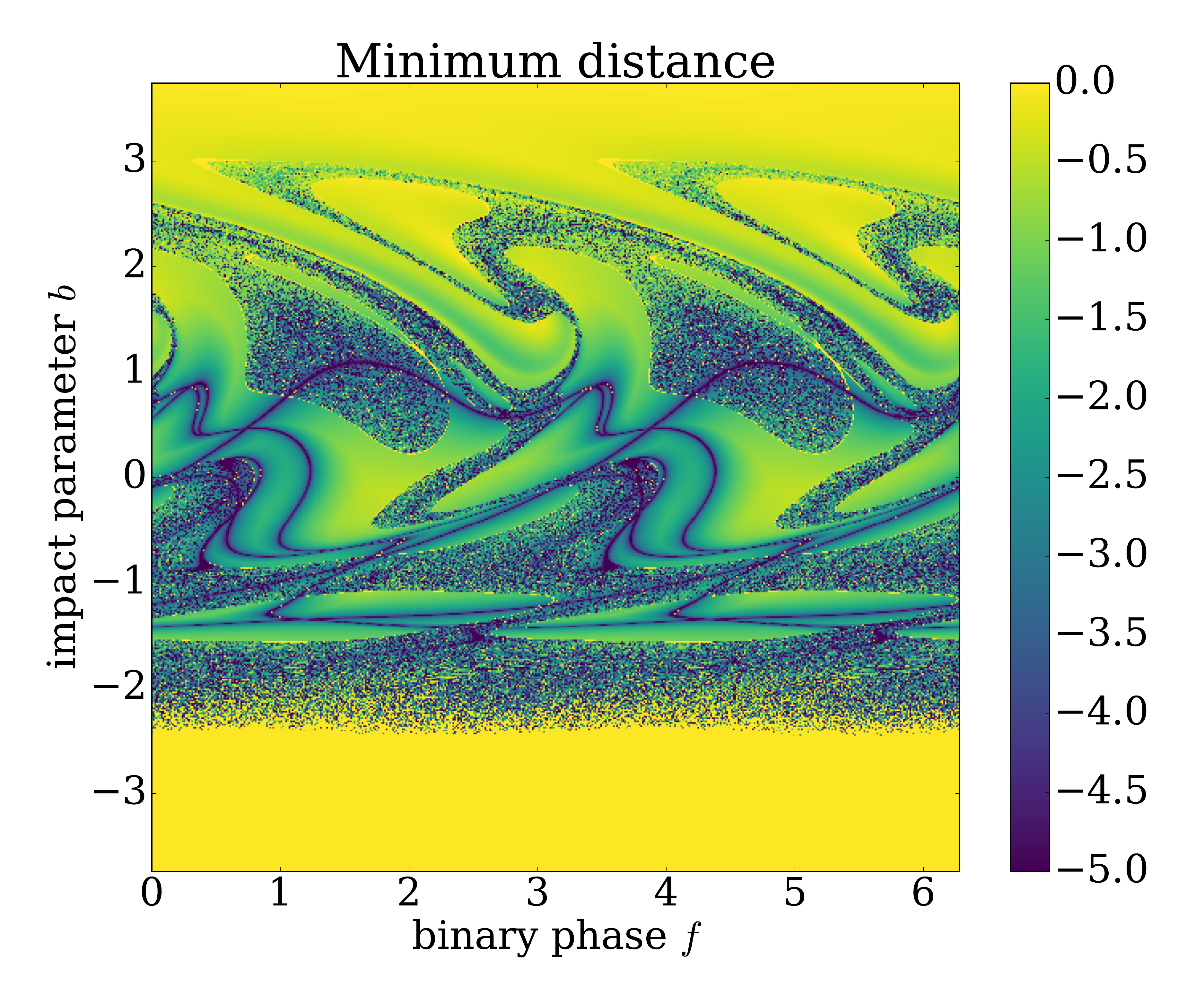}}
\caption{Topological maps derived from binary-single interactions with $a_{0}=1$AU, $v=0.01$, and $\gamma = 0.0$.
\emph{Top left}: Endstates.
\emph{Top right}: Total orbital energy between the endstate binary COM and the remaining single, $\log(E_{\rm BS})$.
\emph{Bottom left}: Number of times the binary-single system splits into an IMS before reaching an endstate, $N_{\rm IMS}$.
\emph{Bottom right}: Minimum distance any two of the three objects undergo throughout an interaction, $R_{\rm min}$, in units of the
initial SMA $a_{0}$, $\log(R_{\rm min}/a_{0})$.
These maps are used to study the classical limit, as described in Section \ref{sec:Classical Limit}.
Further descriptions of outcomes, endstates, and ICs are found in Section \ref{sec:Black Hole Binary-Single Interactions}.
}
\label{fig:classical_outcomes}
\end{figure*}

If the three interacting objects are point-like and only Newtonian gravity is included in the EOM,
the system is said to be scale-free, as the topology and the relative number of each endstate
do not depend on the SMA $a_{0}$ and mass $m$, as long as $v$ remains constant \citep{1983AJ.....88.1549H, Hut:1983js}.
Although such an idealized system does not exist in nature, one approaches this classical scale-free limit as $a_{0}$ increases
relative to the radius of the interacting objects. In this section we study the topology for
$a_{0} = 1$ AU, which provides a reasonable representation of the classical limit.
Results are shown in Figure \ref{fig:classical_outcomes}, and discussed below.

\subsubsection{Endstate Topology}\label{sec:Endstate Topology}

The endstate-map (Figure \ref{fig:classical_outcomes}, upper left) shows a clear
large-scale topological pattern. We find this to emerge from the distribution
of two different small-scale topological pattern types, that we in the following refer to as
the {\it homogeneous-type} and the {\it random-type}. The homogeneous-type is characterized 
by that neighboring points share the same endstate, where the random-type is characterized by
that neighboring points have a semi-random distribution of endstates.
As seen on the figure, the random-type forms one fully connected large-scale structure, which divides the
homogeneous-type into a few disconnected islands. This is a feature of the ultra HB limit for which $v\ll1$, as the random-type structure starts to separate into
isolated parts as $v$ approaches $1$ \citep[e.g.][]{1983AJ.....88.1549H}.

Comparing the $N_{\rm IMS}$-map (Figure \ref{fig:classical_outcomes}, lower left)
with the endstate-map, shows that the random-type
generally originates from RIs and the homogeneous-type from DIs. This is a consequence of the chaotic nature of RIs, which makes the outcomes very sensitive to the ICs. 
Neighboring points on the $f,\ b$ grid in RI regions are
therefore likely to have a semi-random distribution of endstates. In contrast DIs are always prompt. Small variations to the ICs will therefore not have a chance to undergo any significant growth
during this type of interaction. Neighboring points on the $f,\ b$ grid in DI regions therefore often have similar endstates.

The $E_{\rm BS}$-map (Figure \ref{fig:classical_outcomes}, upper right) relates closely to the other maps,
as it is the orbital energy of the single w.r.t. the COM of the first binary that forms (endstate binary or IMS binary),
denoted here by $E_{\rm 1BS}$, that determines if the interaction proceeds as a DI or a RI.
By definition, $E_{\rm 1BS}>0$ leads to DI, in which case $E_{\rm BS} = E_{\rm 1BS}$, where $E_{\rm 1BS}<0$ leads to a RI.
As the distribution of $E_{\rm 1BS}$ seems to vary smoothly across the $f,b$ space, we conclude that
the boarders separating DIs from RIs  --  or equivalently the coastline of the DI islands -- are simply
the path for which $E_{\rm 1BS}=0$. As $E_{\rm 1BS}$ approaches zero along these boarders,
the associated binary-single interaction time correspondingly approaches infinity, which explains why
the `VLint' endstate generally appears between DI and RI regions.

As seen, there is a clear asymmetry in the large-scale topology along the $b$-axis,
despite corresponding symmetry in the positional configuration of the three objects. This asymmetry
results from the way the binary rotates relative to the incoming single,
where $b>0$ and $b<0$ correspond to prograde and retrograde motion, respectively.
For $v\ll1$, the incoming single almost has the same velocity as the binary members as it passes the binary.
In the case of prograde orbits this leads to situations where the three objects almost appear stationary w.r.t. each other.
As a result, for certain combinations of $f,b$ the incoming single does not directly encounter
any of the binary members, which give rise to the large wave-like BS[12] structures seen for $b>0$.
In contrast, negative values of $b$ always result in encounters between the single and a binary member due to the retrograde
motion.  A comparison with the high velocity results shown in  \cite{1983AJ.....88.1549H} confirms that the relative orbital motion indeed is the origin of the $b$-axis asymmetry. For example, the results
in \cite{1983AJ.....88.1549H} clearly show how the topology along the $b$-axis becomes increasingly symmetric as $v$ increases, or in other words, as the velocity of the binary members become less important.

\subsubsection{Kinematical Examples}\label{sec:Kinematical Examples}

Having described a few topological relations in the sections above, we now turn to a more detailed kinematical
analysis. However, we are here only able to provide quantitive descriptions,
as we find that both RIs and DIs generally undergo relative complex orbital evolutions.
We note that an analytical solution for describing the dynamics of the SB limit is possible \citep[e.g.][]{1983ApJ...268..342H},
where a statistical approach has been shown to provide valuable insight in the HB limit \citep[e.g.][]{2014ApJ...784...71S}.
We divide our analysis below into the study of the topology around two orbital configurations: {\it Configuration A} ($f \approx 0$)
and {\it Configuration B} ($f \approx \pi/2$). All results presented below are found by visually inspecting a representative set of three-body interactions.

\begin{figure}
\centering
\includegraphics[width=\columnwidth]{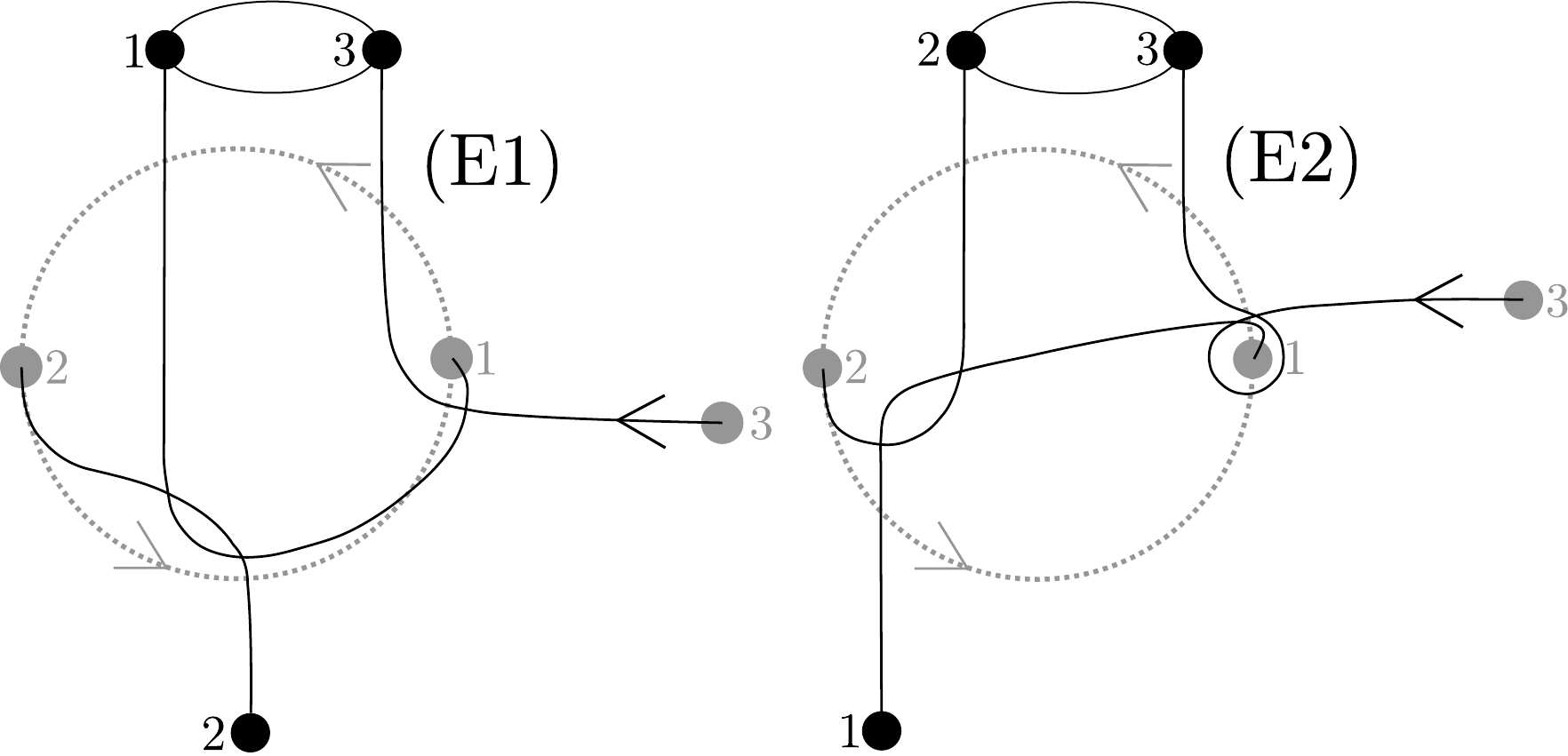}
\caption{Schematic illustrations showing the dynamics of two examples each
representing an exchange interaction through a DI, where the left (E1) leads to BS[13], and the right (E2) to BS[23]. 
The {\it grey large dots} show the three objects at their initial positions,
where the {\it large black dots} show the final configuration. 
The {\it grey dotted circle} illustrates the initial orbit of the two binary members,
`1' and `2', before the incoming object `3' enters the system.
A simplified version of the orbital paths from initial to final state are shown with {\it thin black lines}.
These two examples are further described in Section \ref{sec:Kinematical Examples}.
}
\label{fig:scattering_ill_1}
\end{figure}

\paragraph*{\normalfont Configuration A:}
We first consider the change in topology for varying $b$ at $f \approx 0$.
For this configuration, we refer to the binary member that is closest to the incoming single
as it enters by `1', the remaining member by `2', and thereby the incoming single by `3' (this notation matches the endstate labels on our figures for $f \approx 0 + \pi$).
This configurative labeling is also used in Figure \ref{fig:scattering_ill_1}.

Starting at $b \approx 0$, we see that the topology changes rather abruptly from a BS[23] DI region to a BS[13] DI region.
The kinematics leading to these two neighboring regions is as follows.
For slightly negative values of $b$ object 3 first encounters 1 from below which leads to an upwards deflection of 3
and a downwards deflection of 1. Along its new orbit 1 then interacts with 2, which sends
1 up towards 3 while 2 is being ejected downwards. As a result, [13] escapes as a binary. This interaction
is illustrated to the left (labeled `E1') in Figure \ref{fig:scattering_ill_1}.
For slightly positive $b$ object 3 instead encounters 1 from above, which give rise to an interaction where the
two objects almost perfectly replace each other, which leads to 1 flying to the left and 3 moving upwards. 
Object 1 then interacts with 2 from above, which sends 1 downwards and 2 upwards where it pairs up with 3. As a result, [23] escapes as a binary.
This interaction is illustrated to the right (labeled `E2') in Figure \ref{fig:scattering_ill_1}.

At $b \approx -1.5$ we see a region that is somewhat similar to the one at $b \approx 0$, but with the two
BS[23] and BS[13] regions switched.
For the region at $b \lesssim -1.5$ we find that 3 first interacts with 2 from below after which the binary
has rotated $\approx \pi/2$. This interaction sends 2 downwards and 3 up towards 1. 1 and 3 then undergo a
near parabolic hook-interaction, which sends 3 back towards 2 by ejecting 1 further upwards. As a result, [23] escapes as a binary.
For the region at $b \gtrsim -1.5$ the first interaction between 3 and 2 is simply so strong that 2 is immediately kicked out
by 3 which then takes its place in the binary. As a result, [13] ends up as a binary.

\paragraph*{\normalfont\ Configuration B:} 
We now consider the change in topology for varying $b$ near $f \approx \pi/2$.
For this configuration, we refer to the binary member that moves towards the incoming single by `1',
the remaining member by `2', and thereby the single by `3'. This configuration is similar to that shown in Figure \ref{fig:scattering_ill_1},
but for the binary rotated clockwise by $\pi/2$.

We first notice that for any value of $b \approx 0 \pm 0.5$, the endstate will always be of the same type, in contrast to Configuration A considered above.
The kinematics near $b \approx 0$ are as follows. As object 3 enters the binary, it first accelerates 1 in the upwards direction along its orbital tangent.
Object 3 then interacts with 2 from below, which results in 2 being ejected downwards and 3 being sent up towards 1.
As a result, [13] escapes as a binary. This basically happens for all $b$ for which 3 enters the binary on an orbit that goes in between 1 and 2.

Across larger variations of $b$, significant differences in the topology appear.
For example, interactions near $b \approx 1.5$ all evolve as a RI, whereas interactions
near $b \approx - 1.5$ instead evolve as a DI. We find the kinematical differences to be as follows.
At $b \approx - 1.5$ object 3 no longer moves in between 1 and 2, but instead passes the binary from below. As it passes, 3 and 1 first interact, 
pulling 1 out of its circular motion along an orbit pointing towards the initial motion of 3. Object 3 then interacts with
2 which sends 3 back towards 1 by ejecting 2 in the opposite direction. As a result, [13] escapes as a binary.
At $b \approx 1.5$, the kinematics is similar to that of $b \approx 0$, however, instead or 3 ejecting 2 into an unbound orbit,
2 remains bound after which the interaction then proceeds as a RI.

\subsubsection{Close Encounter Topology}\label{sec:Close Encounter Topology}

The $R_{\rm min}$-map (Figure \ref{fig:classical_outcomes}, lower right), shows a new
structural pattern, that is characterized by connected wavy bands.
These bands map the closest distance any two of the three objects undergo
before the system splits into its first binary-single state (BS[ij] or IMS). 
For example, the band that intersects $f \approx 0,\ b \approx 0$ and rises until
$f \approx \pi/2,\ b \approx 1.0$, originates from the first close encounter between object 1 and 3,
where `1' here refers to the binary member that is closest to the incoming single `3' at $f \approx 0,\ b \approx 0$.
Since the trajectory of 3 is not perturbed significantly
before it first encounters 1, the value of $b$ for which 1 and 3 undergo a close encounter is therefore approximately that of the shadow image
of 1 along an axis parallel to $b$. The first part of the considered band therefore takes the shape of a sinusoidal curve,
which explains why it first rises and then peaks at $f \approx \pi/2$. However, when it goes back towards $b \approx 0$ near $f\approx \pi$
the sinusoidal curve is strongly perturbed, because object 2 comes in between 1 and 3, which leads
to deflections of 3 before it encounters 1.
For velocities $v>1$ object 3 is hardly deflected, and the bands will therefore trace out a near perfect sinusoidal form.

The $R_{\rm min}$-map also shows a population of low $R_{\rm min}$ values inside the RI regions.
This population originates from IMS binaries that are formed with high eccentricity and corresponding
small pericenter distance.

\subsection{Relativistic Limit}\label{sec:Relativistic Limit}

\begin{figure}
\centering
{\includegraphics[width=\columnwidth]{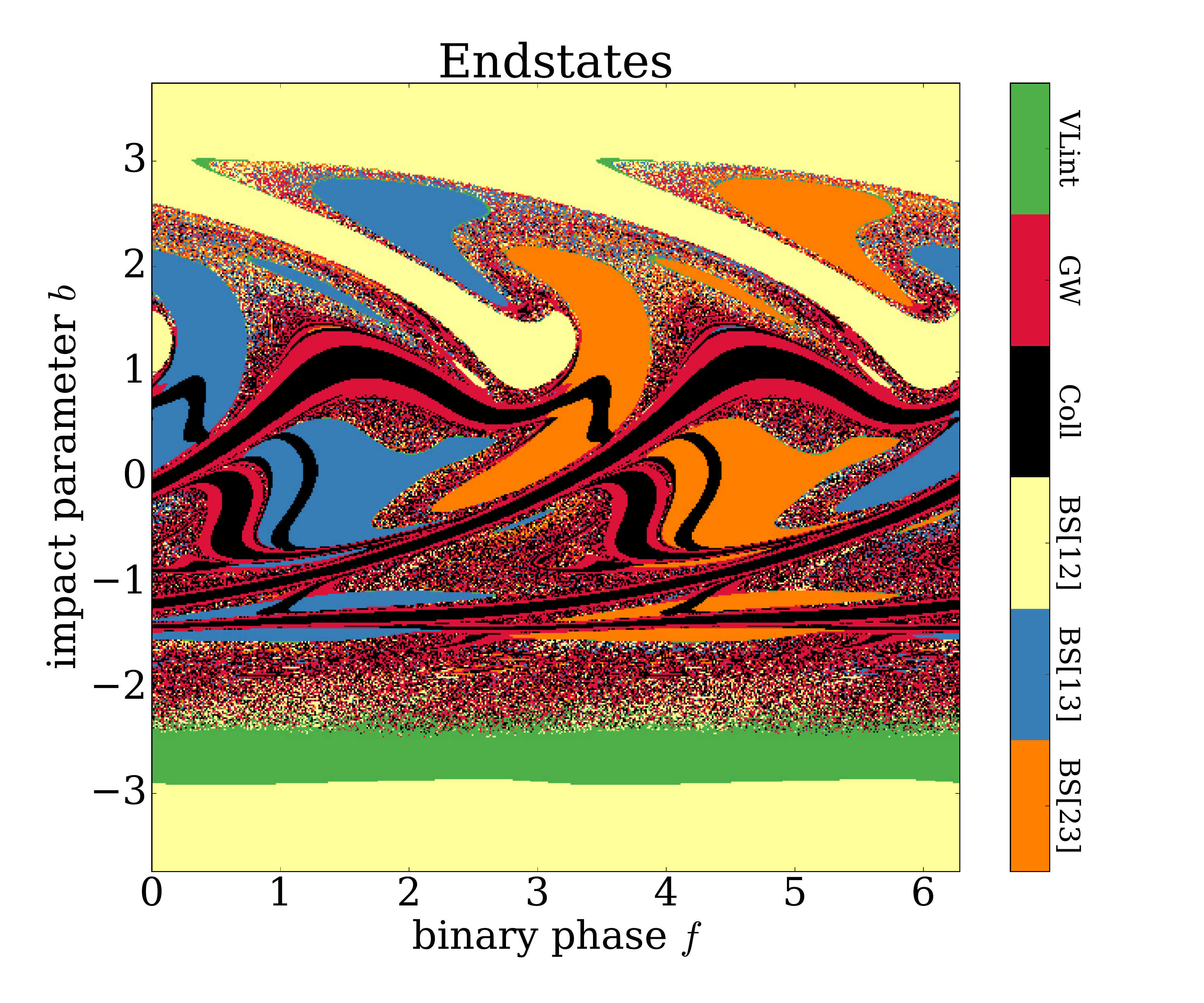}}
{\includegraphics[width=\columnwidth]{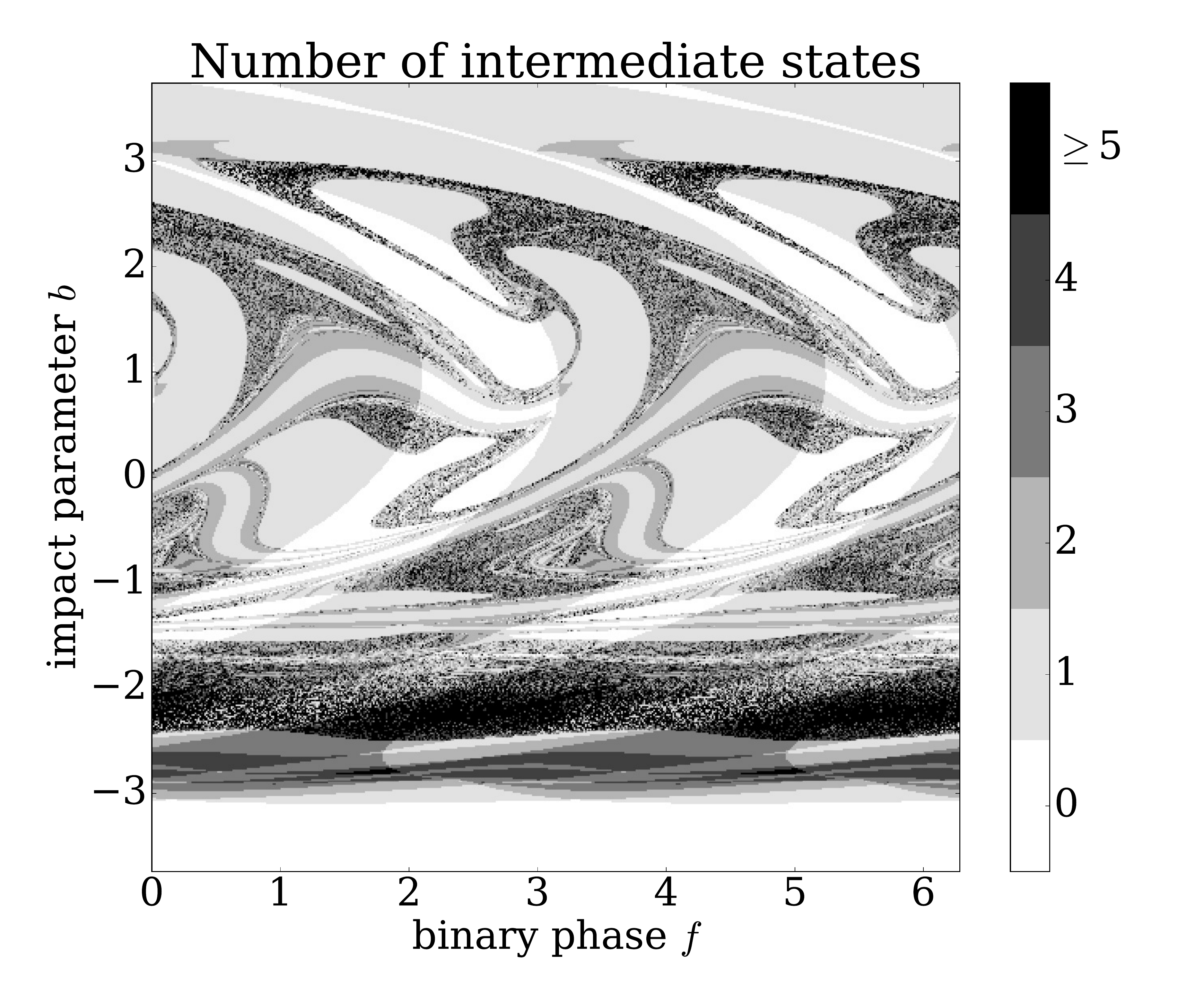}}
\caption{Topological maps derived in a similar way to the ones shown in Figure \ref{fig:classical_outcomes}, but with $a_{0}=10^{-4}$ AU.
\emph{Top plot}: Endstates.
\emph{Bottom plot}: Number of times the binary-single system splits into an IMS before reaching an endstate, $N_{\rm IMS}$.
These maps are used to study the relativistic limit, as described in Section \ref{sec:Relativistic Limit}.
}
\label{fig:relativistic_limit}
\end{figure}

We now move to a study of the topology in the relativistic limit where finite sizes and GR corrections start to play a role. For this, we
perform binary-single scatterings with the initial SMA set to $a_{0} = 10^{-4}$ AU, which naturally leads to a significant
number of very close pairwise BH encounters. This unrealistic low value of $a_{0}$
has only been chosen for illustrative purposes, however, the following results do apply for any $a_{0}$.
Results are shown in Figure \ref{fig:relativistic_limit}, and discussed below.

\subsubsection{Black Hole Collisions}

Looking at the endstate-map (Figure \ref{fig:relativistic_limit}, top plot) and the $N_{\rm IMS}$-map (Figure \ref{fig:relativistic_limit}, bottom plot), we see
that BH collisions form through both DIs and RIs. These two collision populations are described in the following.

The DI-collisions trace the wavy $R_{\rm min}$ bands
shown in Figure \ref{fig:classical_outcomes}, with only small differences due to the modified
dynamics resulting from including GW emission in the EOM. The bands broaden as the value
of $b$ increases. This is due to the corresponding decrease in the relative velocity between the binary member in question and the incoming single,
which leads to an increase in the gravitational focusing, and thereby to the seen broadening in $b$.
As the value of $b$ decreases, the relative velocity increases, which makes the gravitational focusing less efficient. The orbit of the incoming
single therefore has to be relatively fine-tuned to collide with a binary member, which explains the thinner bands at negative $b$.

The RI-collisions originate from the population of
IMS binaries that are formed with a pericenter distance that is smaller than the collisional distance $2R_{\rm s}$.
Despite the apparent randomness of the RI regions, the collision endstates do not seem to be completely randomly distributed in these regions.
For example, the relative number of collisions clearly decreases for $b \gtrsim 2$.
This partially results from the corresponding increase in the total angular momentum,
which naturally leads to fewer low angular momentum orbits.

\begin{figure}
\centering
\includegraphics[width=\columnwidth]{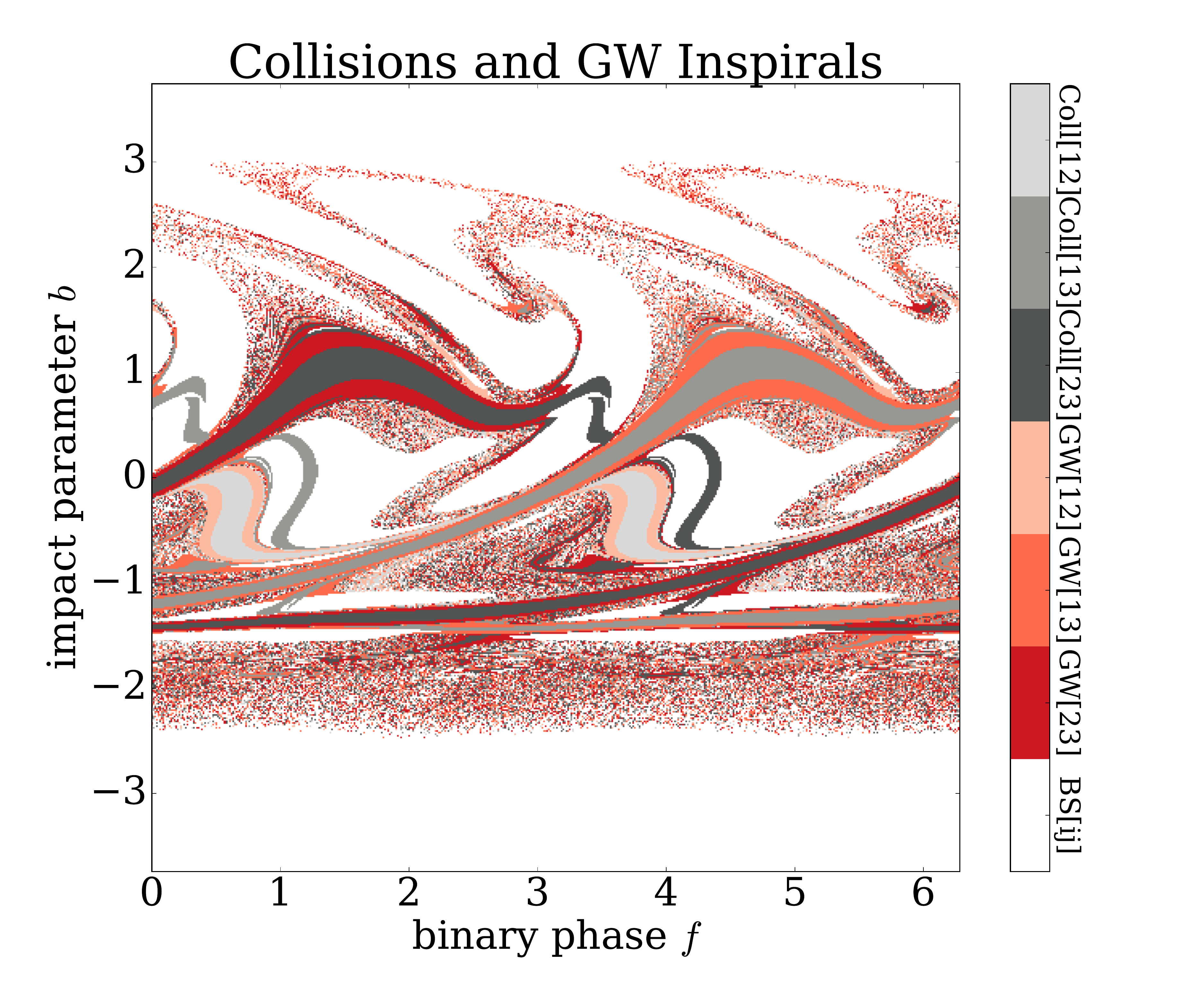}
\caption{Topological map of collision and GW inspiral endstates derived in a similar way to the ones
shown in Figure \ref{fig:classical_outcomes}, but with $a_{0}=10^{-4}$ AU.
This map provides information about what object pair that collides or undergoes a GW inspiral.
As seen, the {\it red colors} show GW inspirals, where the {\it grey colors} show collisions.
This map is discussed in Section \ref{sec:Relativistic Limit}. 
}
\label{fig:collGW_map}
\end{figure}

Further insight into the collision endstates distribution is provided in Figure \ref{fig:collGW_map}, which
shows what pair that collides. Besides bands originating from collisions with the incoming
single (Coll[13], Coll[23]), we interestingly see a band passing through $f \approx 0.7,\ b \approx -0.5$ that is formed by collisions of the binary members themselves (Coll[12]).
We find that the corresponding
dynamics is such that the incoming single, during its first passage, simply perturbs the
two binary members into a prompt collisional orbit. Although this seems like a fine-tuned problem, the corresponding bands appear
relative broad, which indicates a non-negligible formation probability.

\subsubsection{GW Inspirals}

A clear population of GW inspirals is seen near the boarders of the
wavy collision bands. The origin of this inspiral population is not surprising,
as the collision boarders are characterized by near collisional orbits. When GW
emission is included in the EOM, these orbits therefore naturally result in strong
GW emission, which then leads to the observed populations of GW inspirals.
Just as the population of collisions in the wavy bands forms through DIs, so do the GW inspirals forming along the bands,
as seen in the bottom plot in Figure \ref{fig:relativistic_limit}.

We note that there are a few collisional bands located in the DI regions with no GW inspirals forming along
their edges. We find here that the dynamics leading to the prompt collision at the same time
ejects the third BH into an unbound orbit. The endstates populating these bands are therefore characterized by a collision with
an unbound companion. This explains why no GW inspirals form along their edges, as the binary-single system splits up into a BS[i,j] state
before the GW inspiral is complete. The two binary objects will of course still undergo the GW inspiral, but it will happen without a bound companion.

Moving away from the wave-like bands, we see how GW inspirals clearly form throughout the RI regions.
We note here that collisions and GW inspirals do not have to follow the same distribution, as the 
formation of a GW inspiral is not related to a particular value of the IMS binary pericenter distance unlike a collision. Instead, the formation of a
GW inspiral relates to how much energy it can lose through GW emission during the orbital time it is isolated from the bound single \citep{2014ApJ...784...71S}. 
As a result, GW inspirals can form from IMS binaries with both small and large pericenter distances, depending on the corresponding
isolation time, and can therefore appear in regions where collisions are relative rare.
For example, in the upper plot in Figure \ref{fig:relativistic_limit} there are many GW inspirals forming for $b > 2.5$ but only relative few collisions. 
A more detailed picture is provided in Figure \ref{fig:collGW_map}, which also shows what pair that undergoes the GW inspiral.

\subsection{Out-of-Plane Interactions}

\begin{figure}
\centering
\includegraphics[width=\columnwidth]{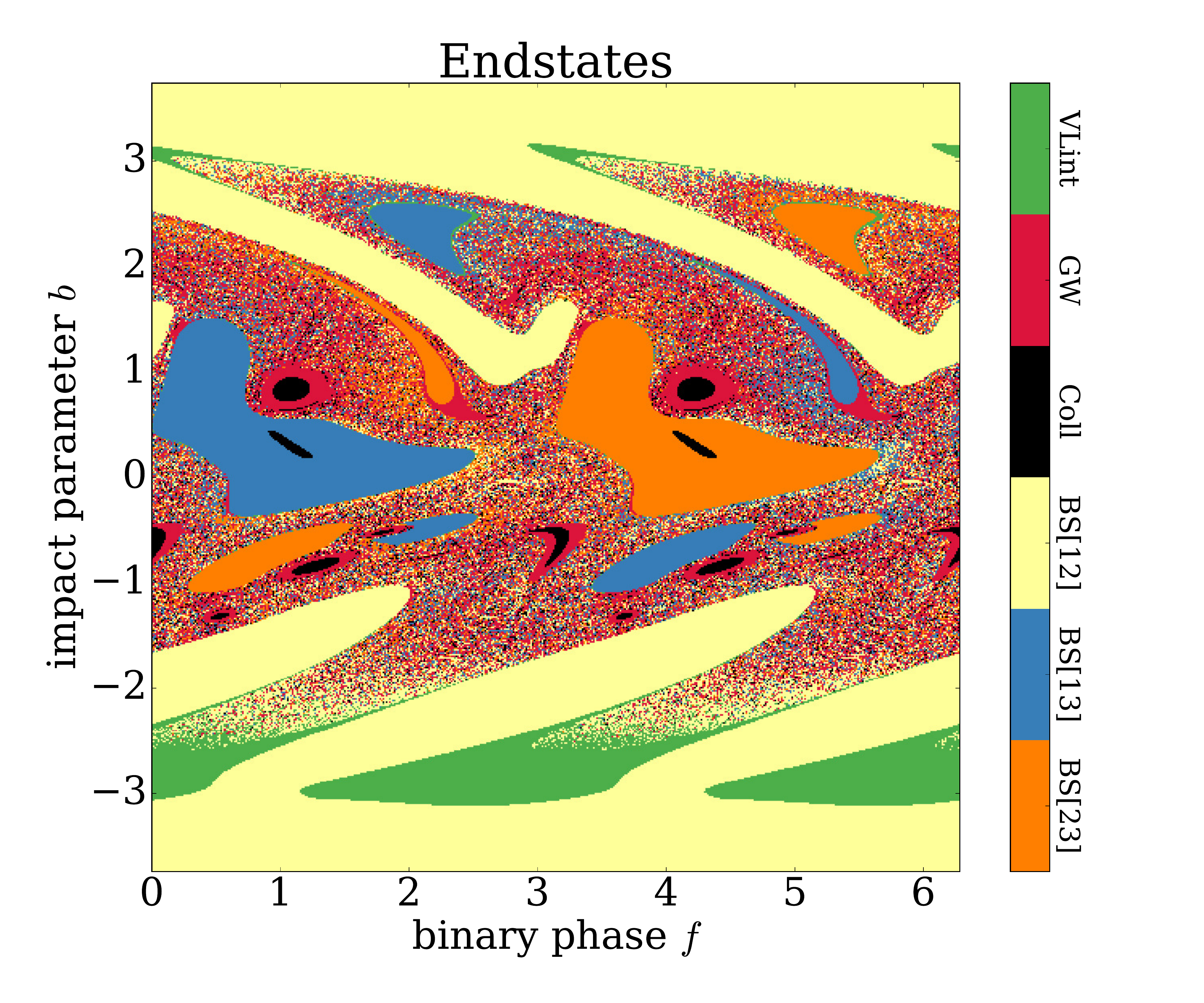}
\includegraphics[width=\columnwidth]{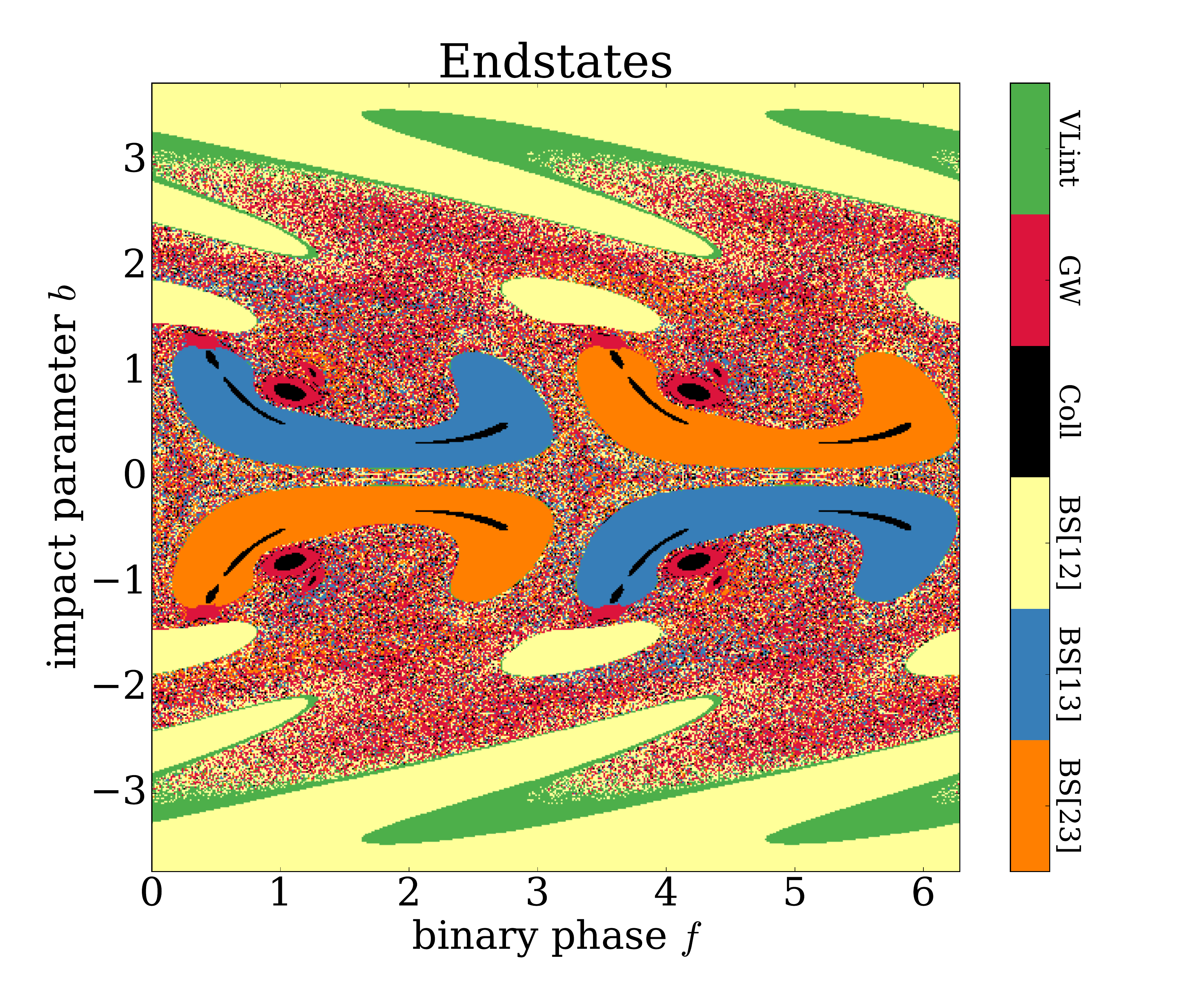}
\caption{Topological endstate-maps derived in a similar way to the ones shown in Figure \ref{fig:classical_outcomes}, but with $a_{0}=10^{-4}$ AU
and varying $\gamma$.
\emph{Top plot}: Endstate topology for $\gamma = \pi/4$.
\emph{Bottom plot}: Endstate topology for $\gamma = \pi/2$.
The corresponding endstate topology for $\gamma = 0$ is shown in the top of Figure \ref{fig:relativistic_limit}.
}
\label{fig:out_of_plane_int}
\end{figure}

For completeness, we here study the change in topology for varying orbital-plane angle $\gamma$.
Endstate-maps for $\gamma = \pi/4$ and $\gamma = \pi/2$ are shown in
Figure \ref{fig:out_of_plane_int}, and briefly discussed below.

The most notable difference between the co-planar case ($\gamma = 0$)
and our presented out-of-plane cases ($\gamma  = \pi/4,\ \pi/2$), is a breakup of the connected
wave-like collision and GW inspiral bands seen for $\gamma = 0$. To understand this change, we note that the
incoming single can only undergo a prompt collision or a prompt GW inspiral near regions where its orbital plane
intersects with the binary orbit. When $\gamma = 0$, the incoming single
intersects for sinusoidal variations in $f,b$, however, when $\gamma > 0$, the orbital plane of the
single is only able to intersect the binary orbit at maximum two points, namely at $\{ -\pi/2,+\pi/2\}$. This correspondingly result in
isolated regions in the topological map instead of connected bands.

As with the co-planar case, the DI collision regions for $\gamma > 0$
are surrounded by GW inspirals that form through a prompt capture. However, a large population of both DI and RI
collisions and GW inspirals also form beyond these DI regions. In fact, we find that the total number of collisions and GW inspirals is not strongly
dependent on $\gamma$, although the distribution of orbital parameters at merger might depend on $\gamma$ due to 
the corresponding change of angular momentum. We save such a study for future work.

\section{The Role of GW emission}\label{sec:The Role of GW radiation}

We now perform a slightly more controlled study on how the inclusion of GW emission in the EOM
affects the topology on different scales. For this, we explore two different topological regions, with and without GW emission
included in the EOM. For all interactions we set $a_{0} = 10^{-4}$ AU and $\gamma = 0$ to have GW emission effects
stand out clearly. Results are discussed below.

\subsection{Large-Scale Effects}\label{sec:Large Scale Effects}

\begin{figure}
\centering
\includegraphics[width=\columnwidth]{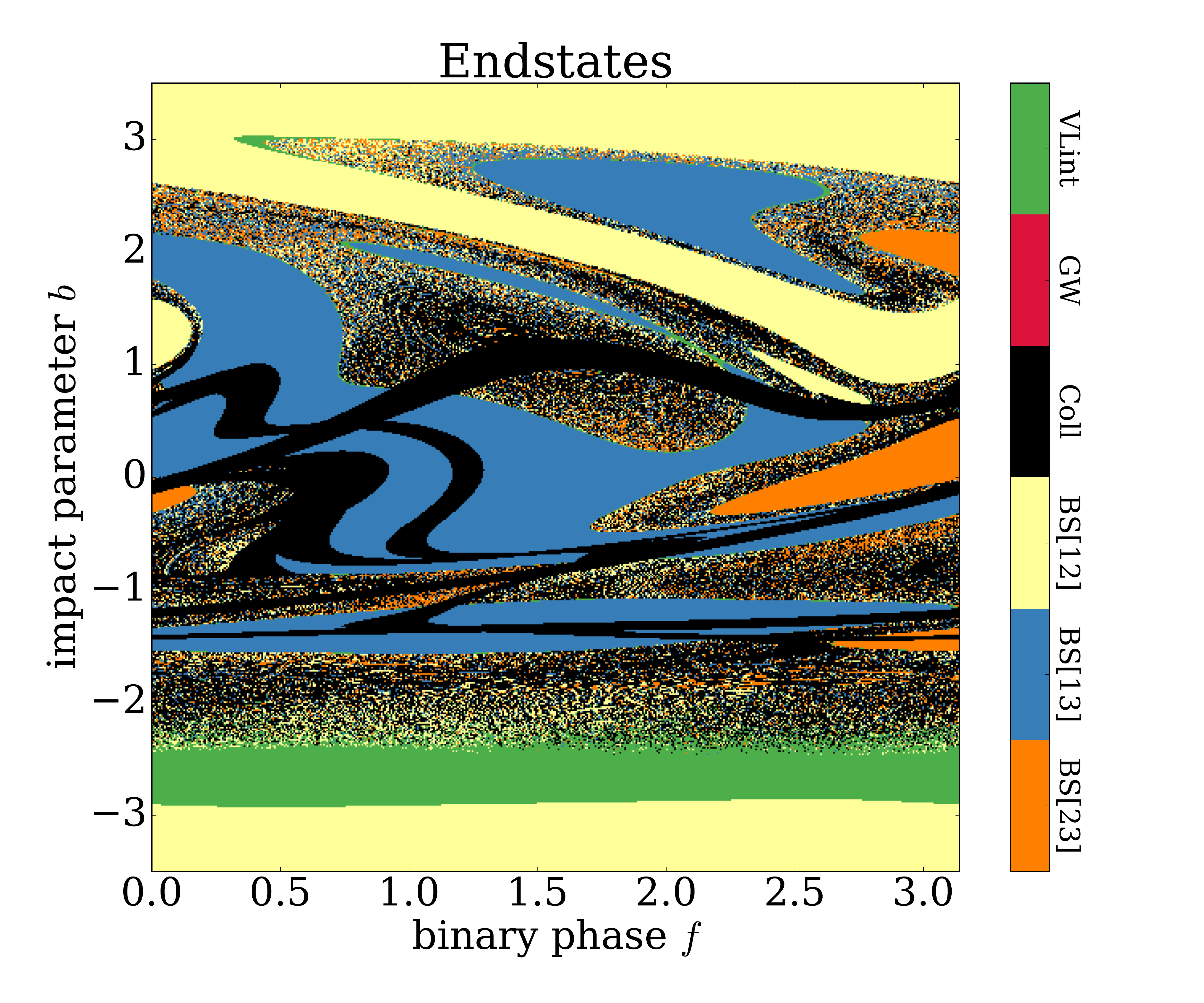}
\includegraphics[width=\columnwidth]{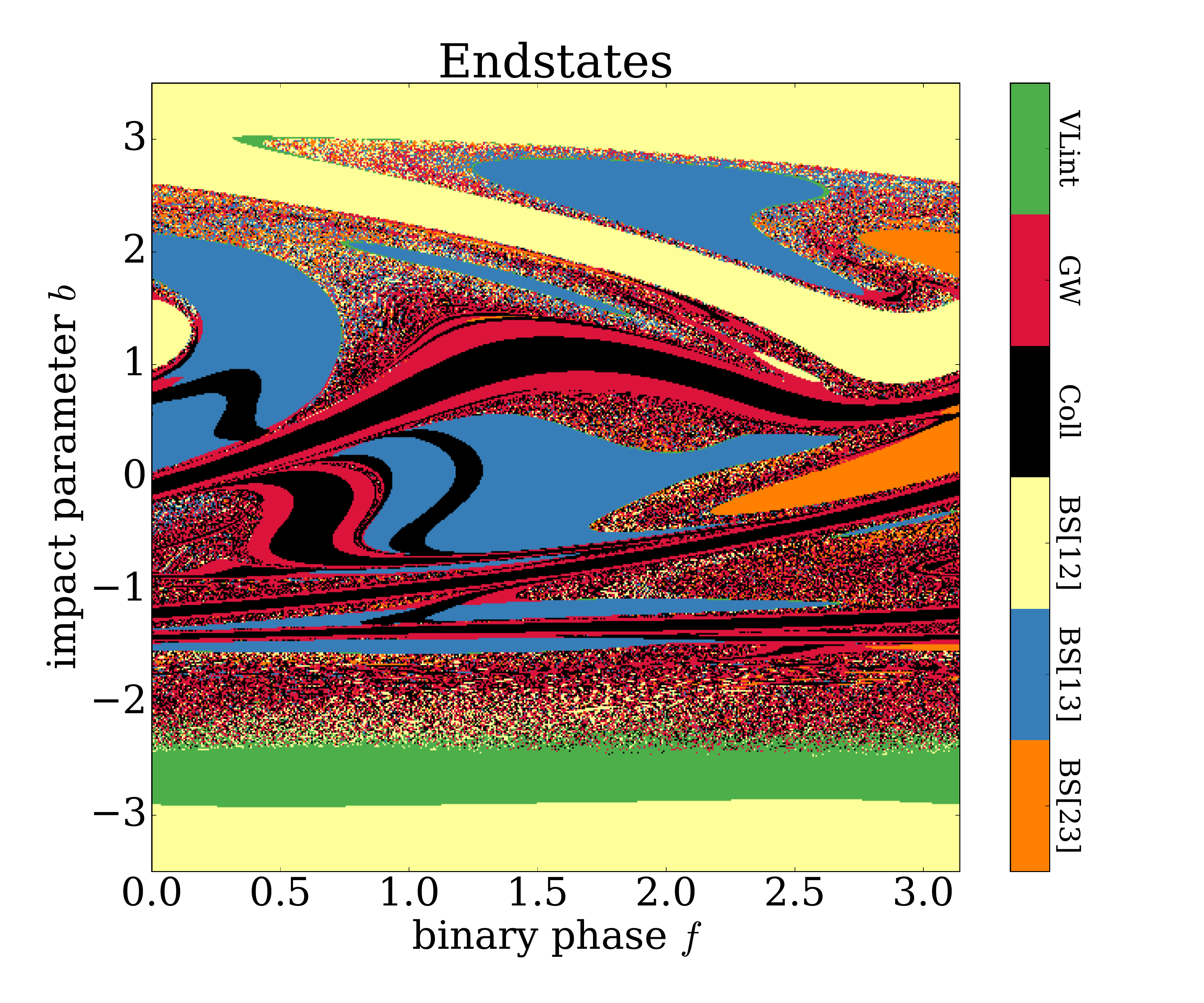}
\caption{Topological endstate-maps derived in a similar way to the ones shown in Figure \ref{fig:classical_outcomes}, but with $a_{0}=10^{-4}$ AU.
\emph{Top plot}: Endstate-map derived {without} the inclusion of GW emission in the $N$-body EOM.
\emph{Bottom plot}: Endstate-map derived {with} the inclusion of GW emission in the $N$-body EOM.
Results are discussed in Section \ref{sec:Large Scale Effects}.}
\label{fig:W_WO_GW_largescales}
\end{figure}

We start by analyzing how the inclusion of GW emission affects the
large-scale topology. Results are shown Figure \ref{fig:W_WO_GW_largescales}.

When GW emission is not included in the EOM (top plot), we see that the collision regions simply appear superimposed on
the classical endstate-map, with a distribution that follows the $R_{\rm min}$ distribution as expected. Other than occupying part of the map,
the inclusion of collisions does not affect the large-scale topological structure, as the EOM is unchanged.

When GW emission is included (bottom plot), we see that this not only leads to the formation of GW inspirals,
but also a slight modification of the topology. For example, at $f \approx 0.75$, $b \approx 0.75$, where the collision band intersects
the RI region in the classical map, the inclusion of GW emission clearly smoothes
out the transition region, by turning a part of the BS[1,3] endstates formed in the DI region into a semi-random mix of outcomes.
Generally, one finds that the largest topological changes indeed are found near regions where collision endstates,
DIs, and RIs intersect. This results from two factors: the GW emission is strong (the objects are close to colliding) and the
the orbital energy $E_{\rm 1BS}$ is near zero (the boarder between a DI and a RI is characterized by $E_{\rm 1BS} = 0$).
This combination makes it possible for the GW emission to change a classical DI into a RI, by emitting the orbital energy out of the system that otherwise would
have ejected the single into an unbound orbit. In fact, in \cite{2016arXiv160909114S} it was briefly discussed if this way of  `capturing' the single into a RI through
GW emission would be able to change the relative frequency of endstates. We have here shown and argued that this indeed happens,
but only in a very small part of the available phase space. The effect is therefore negligible from a statistical perspective.

\begin{figure}
\centering
\includegraphics[width=\columnwidth]{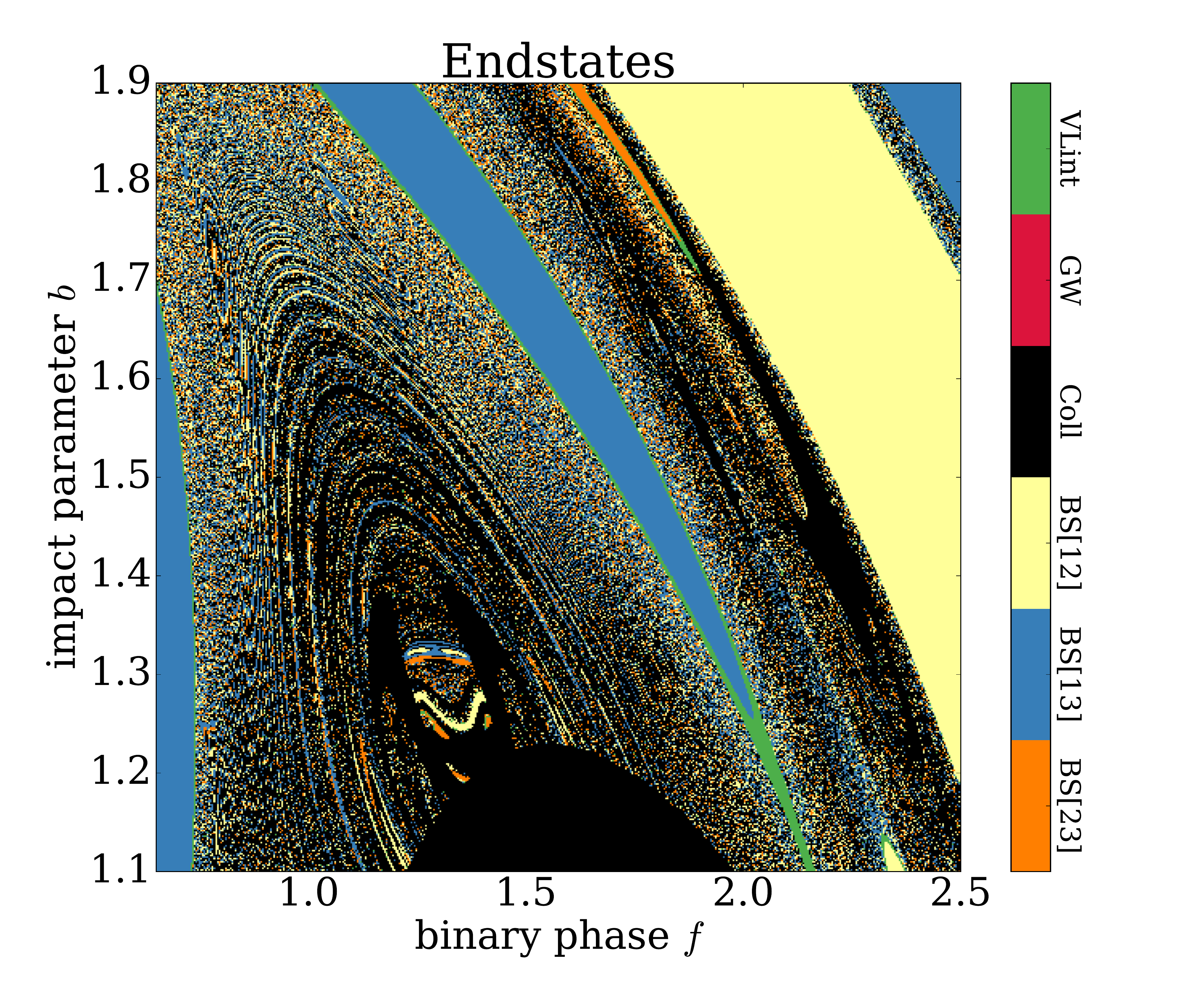}
\includegraphics[width=\columnwidth]{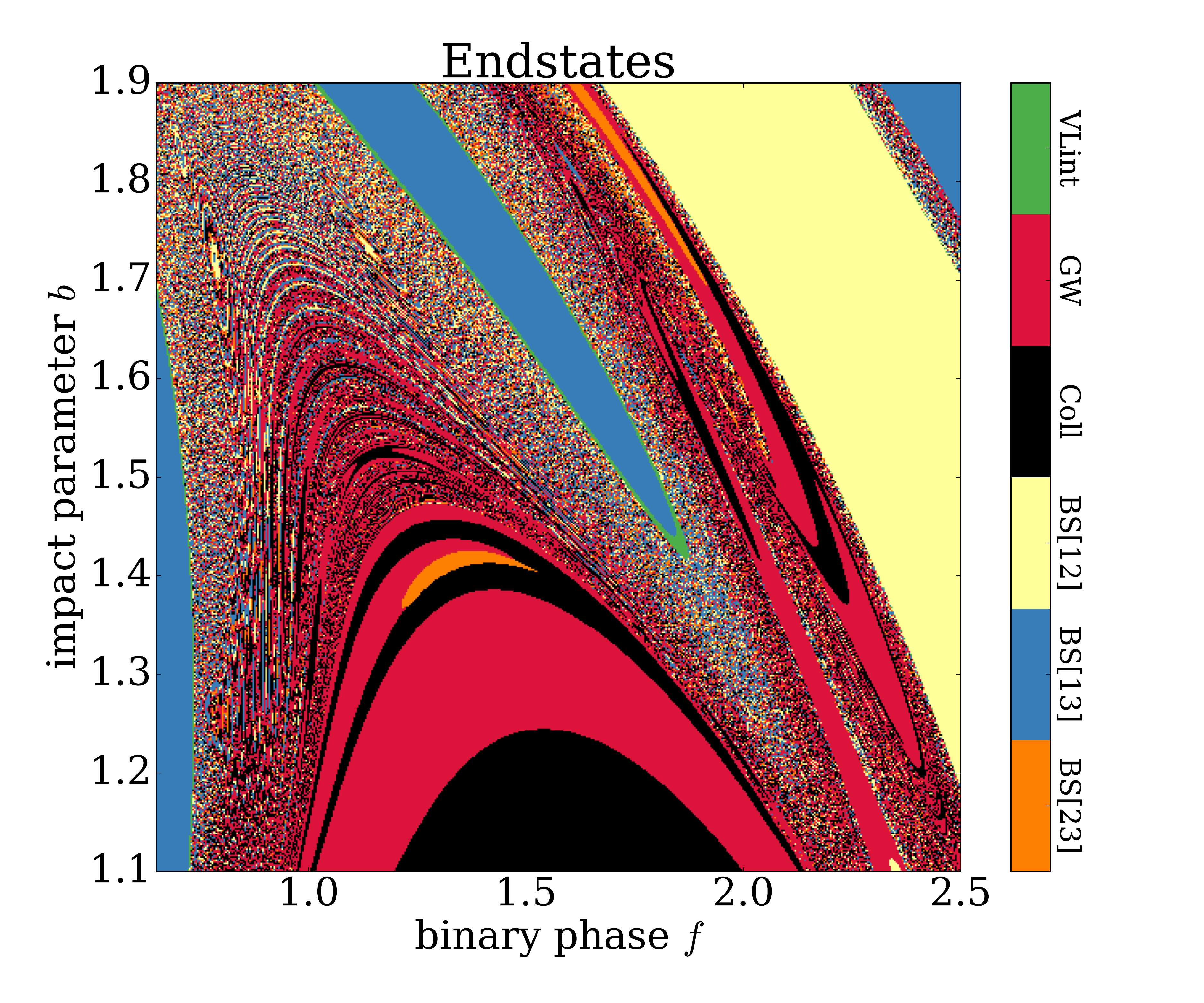}
\caption{Topological endstate-maps derived in a similar way to the ones shown in Figure \ref{fig:classical_outcomes}, but with $a_{0}=10^{-4}$ AU.
\emph{Top plot}: Endstate-map derived {without} the inclusion of GW emission in the $N$-body EOM.
\emph{Bottom plot}: Endstate-map derived {with} the inclusion of GW emission in the $N$-body EOM.
Results are discussed in Section \ref{sec:Small Scale Effects}.}
\label{fig:W_WO_GW_smallscales}
\end{figure}

\subsection{Small-Scale Effects}\label{sec:Small Scale Effects}

We now study how the topology changes on small scales
as GW emission is included in the EOM. Results are shown in Figure \ref{fig:W_WO_GW_smallscales}.

In the top plot, where GW emission is not included, we first notice that the considered RI region
has a surprisingly rich and complex micro-topology. This will be studied in greater detail in Section \ref{sec:Topological Microstructures}.
As in the large-scale example, we see here how the collision regions simple appear as sharp shadows superimposed on top of the classical
map. 

We now turn to the bottom plot where GW emission is included. As seen, the inclusion drastically changes the topology close to
the classical collision bands. Not only is a clear population of GW inspirals forming along the bands, but the
GW emission further results in band-like structures that alter between collisions and GW inspirals.
To understand the kinematics leading to this alternation we visually inspected a set of binary-single interactions
for $f\approx \pi/2$ and varying $b$. For this configuration we here denote
the binary member moving away from the incoming single by `2', the remaining member
by `1', and thereby the incoming single by `3'. We find the kinematics to be as follows.
First, the large collisional band at the bottom ($b \approx 1.15$) is formed by 2 directly colliding with 3 as it enters the binary, as also described
in Section \ref{sec:Close Encounter Topology}.
Above this is an inspiral band ($b \approx 1.3$), which we find consists of interactions where 3 instead of directly colliding with
2, slightly grazes it, which then leads to a prompt GW inspiral between the two.
The subsequent collision band ($b \approx 1.4$) is formed by 3 first being captured into a relative wide
low angular momentum orbit by 2, after which 3 and 2 collide at their first pericenter passage along their new orbit. The following narrow BS[23] band ($b \approx 1.42$)
arises from a rather fine-tuned dynamical configuration where 1 becomes kinematically unbound just before 2 and 3 undergo a collision or a GW inspiral. This band
is therefore sensitive to the tidal threshold used for labeling a BS[ij] endstate, and is therefore not a fundamental feature.
The next few bands form in a similar way to the previous collision
band, only with 2 and 3 changing between undergoing a collision or a GW inspiral.
To conclude, these topological changes all relate to the capture of object 3 into a bound orbit partly through GW emission.
As the capture happens at low angular momentum, and angular momentum further is radiated away during the capture, the system
evolves in a state that is likely to result in either a prompt collision or a GW inspiral.

\section{Topological Microstructures}\label{sec:Topological Microstructures}

\begin{figure}
\centering
\includegraphics[width=\columnwidth]{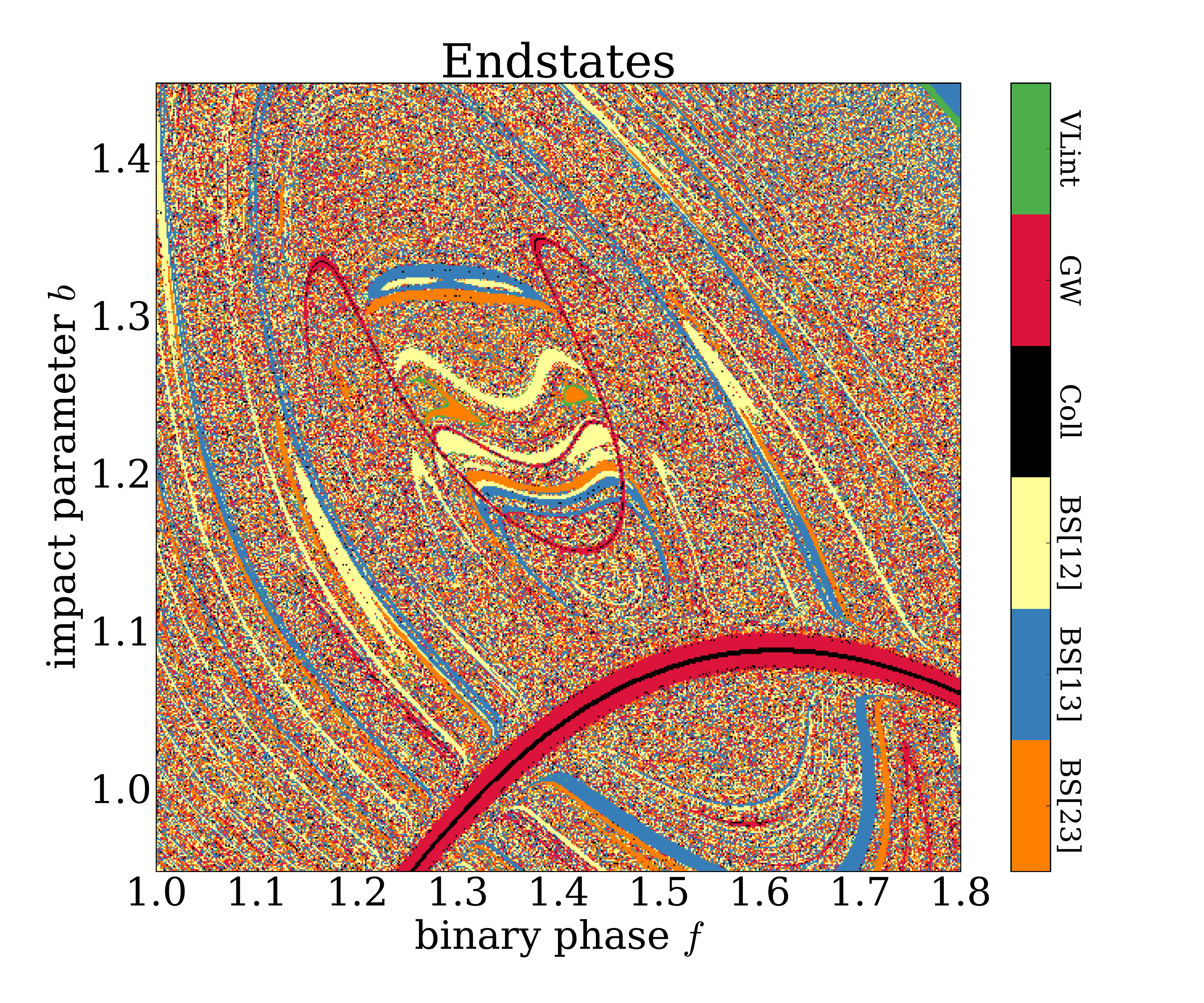}
\includegraphics[width=\columnwidth]{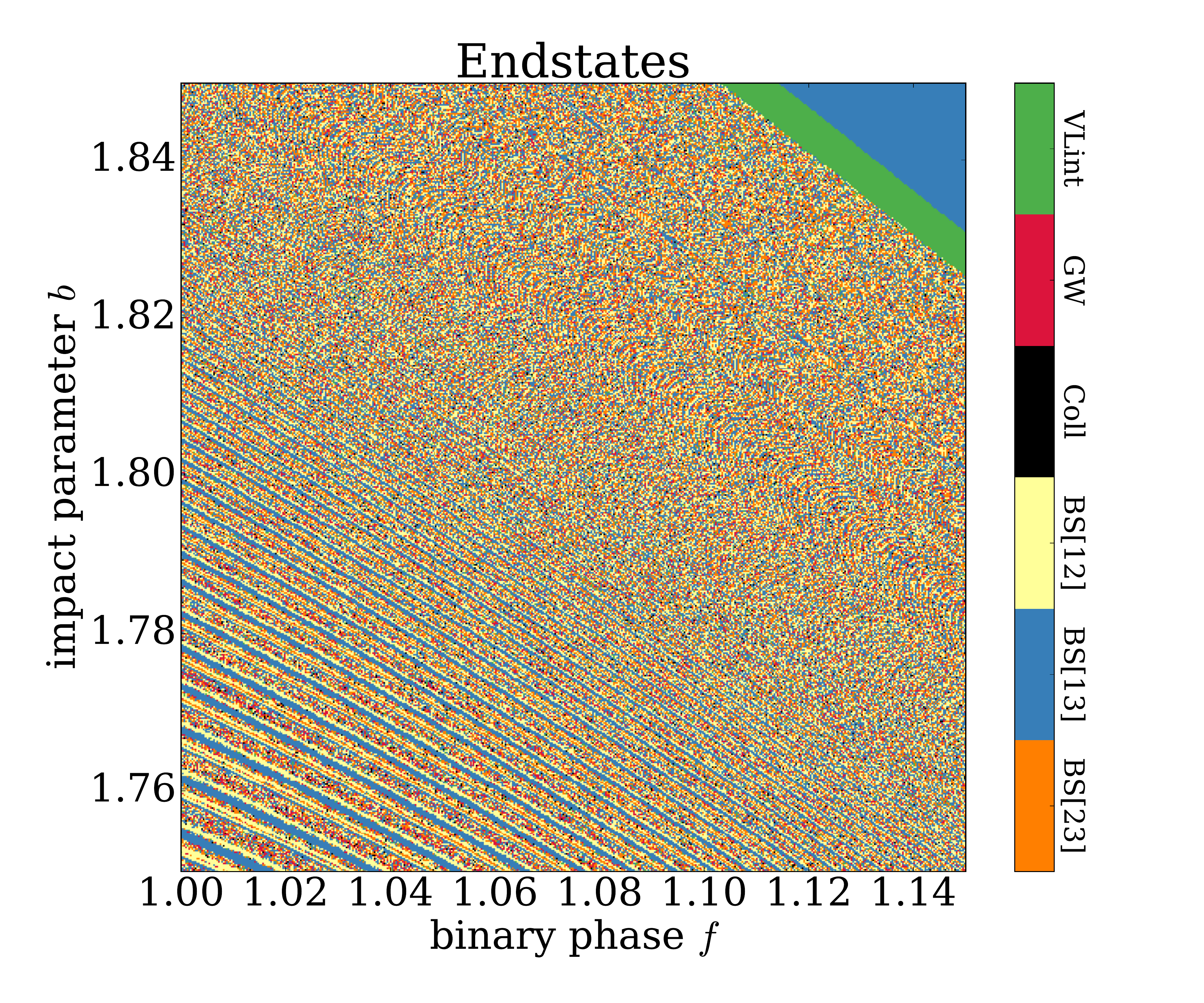}
\caption{Topological endstate-maps derived in a similar way to the ones shown in Figure \ref{fig:classical_outcomes} with $a_{0} = 1$ AU and $\gamma = 0$.
{\it Top plot}: Zoom in on the center of one of the larger RI regions.
{\it Bottom plot}: Zoom in on the boundary between one of the larger RI regions and its neighboring DI region.
As seen in both plots, the considered regions do not just consist of a semi-random distribution of endstates as it appears
to on larger scales. Instead, the regions have in fact a rich variety of micro-topological structures.
The origin of these structures is discussed in Section \ref{sec:Topological Microstructures}.
}
\label{fig:micro_top_struc}
\end{figure}

\begin{figure}
\centering
\includegraphics[width=\columnwidth]{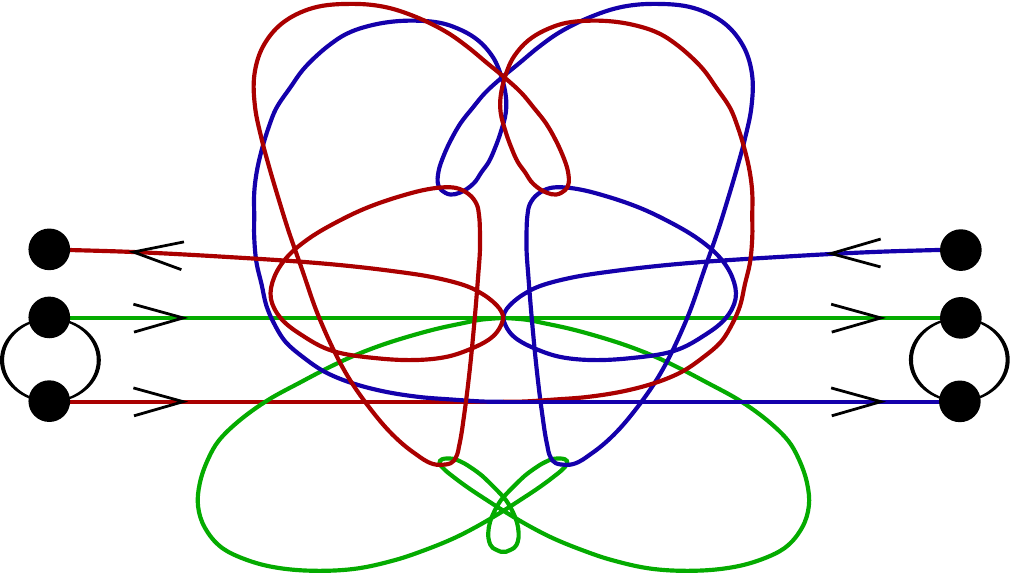}
\caption{Schematic illustration of the orbital trajectories associated with an interaction leading to a BS[23] endstate
found near $f \approx 1.3,\ b \approx 1.25$.
As seen, the objects are in this interaction clearly moving in a non-chaotic way on orbits that trace out an elegant symmetrical pattern. 
We find that most of the interactions near the considered values for $f,\ b$ actually go through this orbital evolution, but with different outcomes
to follow depending on the exact ICs. In this illustration the colors {\it red}, {\it green}, and {\it blue} refer to object $1$, $2$, and $3$, respectively,
where the arrows denote the direction of motion. The considered BS[23] outcome corresponds to the case where the binary enters from the left
and moves towards the right. Further discussions are found in Section \ref{sec:Zoom Region 1}.}
\label{fig:orbit_ex_1}
\end{figure}

We now explore the topology of the microstructures that surprisingly seem to appear within RI regions.
For this, we perform scatterings for $a_{0} = 1$ AU, and $\gamma = 0$, with
collisions and GW emission included in the EOM.
However, we do note that our considered small scale topology is likely to change slightly if additional PN terms are added to the EOM,
we therefore limit ourself to general conclusions. Results from two selected zoom in regions, referred to as `Zoom Region 1' and `Zoom Region 2', respectively,
are shown in Figure \ref{fig:micro_top_struc}, and discussed below.

\subsection{Zoom Region 1}\label{sec:Zoom Region 1}

The top plot shows a zoom in on one of the larger RI regions. As seen, the region does not only consist of
semi-randomly distributed endstates, as it appears to on large scales, but in fact hosts clear micro-topological structures of various kinds.

To gain insight into the dynamics of this region, we visually investigated a representative set of binary-single interactions.
From this exercise we found that most of the interactions near the center ($f \approx 1.3,\ b\approx 1.2$) start out in a similar way,
where the objects first interact non-chaotically with their orbits tracing out a well defined pattern with a clear degree of symmetry.
A schematic illustration of this primary orbital interaction assuming perfect symmetry is shown in Figure \ref{fig:orbit_ex_1}.
Although the interaction in this example promptly ends as a BS[23] endstate, the outcomes do vary depending on the exact ICs. In a few
cases we even observed that the system after this primary interaction entered a quasi-stable state which it stayed in for 5-10 cycles. We note here
that the more stable a three-body state is, the less likely it is to form \citep[e.g.][]{1983AJ.....88.1549H}, which implies
that fully stable three-body states are not assembled through our binary-single experiments (Newtonian three-body states
can not be stable at plus infinity if they are not stable at minus infinity).
Of course the inclusion of GR in the EOM immediately changes the situation, as the implied GW emission destabilizes the system
within a finite time. Further studies on the possible assembly of quasi-stable three-body states with and without GR is beyond this paper,
but still of great theoretical interest \citep[e.g.][]{2013PhRvL.110k4301S}.

Although the state illustrated in Figure \ref{fig:orbit_ex_1} is not stable, it certainly classifies as a special interaction due to its
symmetry and interesting orbital topology. However, the classification of such interactions is not entirely clear and to our knowledge
still unexplored. A systematic search through the $f,\ b$ space could reveal if this is a member of
a larger family of peculiar interactions, or just a single interesting incident.

\subsection{Zoom Region 2}

The bottom plot shows a zoom in on a transition zone between RIs (lower left) and DIs (upper right). As seen, the only
pattern that is clearly visible consists of nested rings with a spacing that decreases as
one moves closer to the DI boundary from the RI side. 
Focusing on the BS[13] endstate bands, we find that the corresponding IMS binary configuration is such
that when the single returns from its first excursion, the resultant IMS binary-single interaction
leads to a prompt BS[13] endstate. This happens in each band, with the only difference being how many times the
IMS binary rotates before the single returns. As a result, there is a band for each
integer $n$, where $n$ is the number of times the IMS rotates before the single
returns (by visual expectation we find that the band in the lower left corner corresponds to $n=10$).
As $E_{\rm 1BS}$ approaches zero as one moves towards the DI region, the number $n$ will correspondingly
approach infinity. As a result, the spacing between the blue bands becomes smaller and smaller,
approaching zero near the boundary. This explains why substructures with infinitely small spacings exist and where they are located.
The precess of such structures implies that the true richness of micro topological structures cannot be studied with a finite grid approach
as the one we have presented here.

\section{Double Gravitational Wave Mergers}\label{sec:Double GW Inspirals}

All GW inspirals form during the binary-single interaction while the three objects are still bound to each other in an IMS \citep{2014ApJ...784...71S}.
Ignoring dynamical kicks caused by asymmetric GW emission \citep[e.g.][]{2006ApJ...653L..93B}, this interestingly implies that the BH formed from
the GW inspiral merger and the remaining single are still bound to each other after the three-body interaction. A compelling question is how long time
it takes for this binary to merge through GW emission, or equivalently, what the time span is from the first to the second GW merger.
If the time span is short enough, it might be possible for LIGO to observe both GW mergers, a scenario that undoubtful would be extremely exiting.

In this section we explore for which combinations of $f$ and $b$ the time span from first to second GW merger, denoted by ${\Delta}t_{12}$,
takes its shortest value. A more detailed study of this double GW merger scenario is reserved for a separate paper, and the following
discussions are therefore kept short. An illustration of our proposed double GW merger channel is shown in Figure \ref{fig:double_merger_ill},
and results presented below.

\begin{figure}
\centering
\includegraphics[width=\columnwidth]{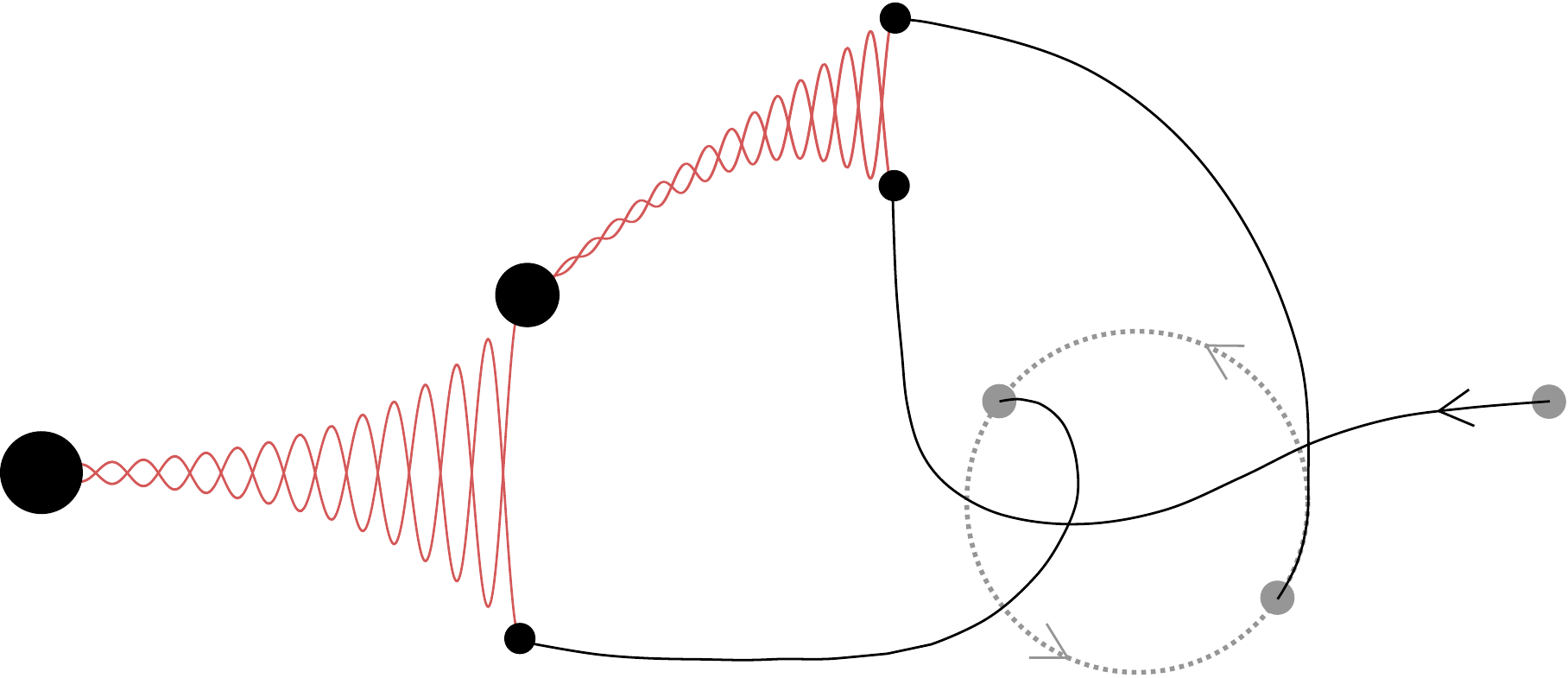}
\caption{Illustration of a binary-single interaction that results in the formation of two GW mergers.
As seen, the interaction first results in the formation of a BBH that promptly undergoes a GW inspiral (first GW merger).
After this, the resultant BH formed through the first GW inspiral then subsequently
undergoes a GW merger with the remaining bound single (second GW merger).
If the time span between the first and the second GW merger is relative short, then it might be possible to
observe both GW mergers with a detector such as LIGO. In Section \ref{sec:Time from First to Second GW Merger}
we explore which values for $f,\ b$ that lead to the shortest time span.}
\label{fig:double_merger_ill}
\end{figure}

\subsection{Time from First to Second GW Merger}\label{sec:Time from First to Second GW Merger}

\begin{figure}
\centering
\includegraphics[width=\columnwidth]{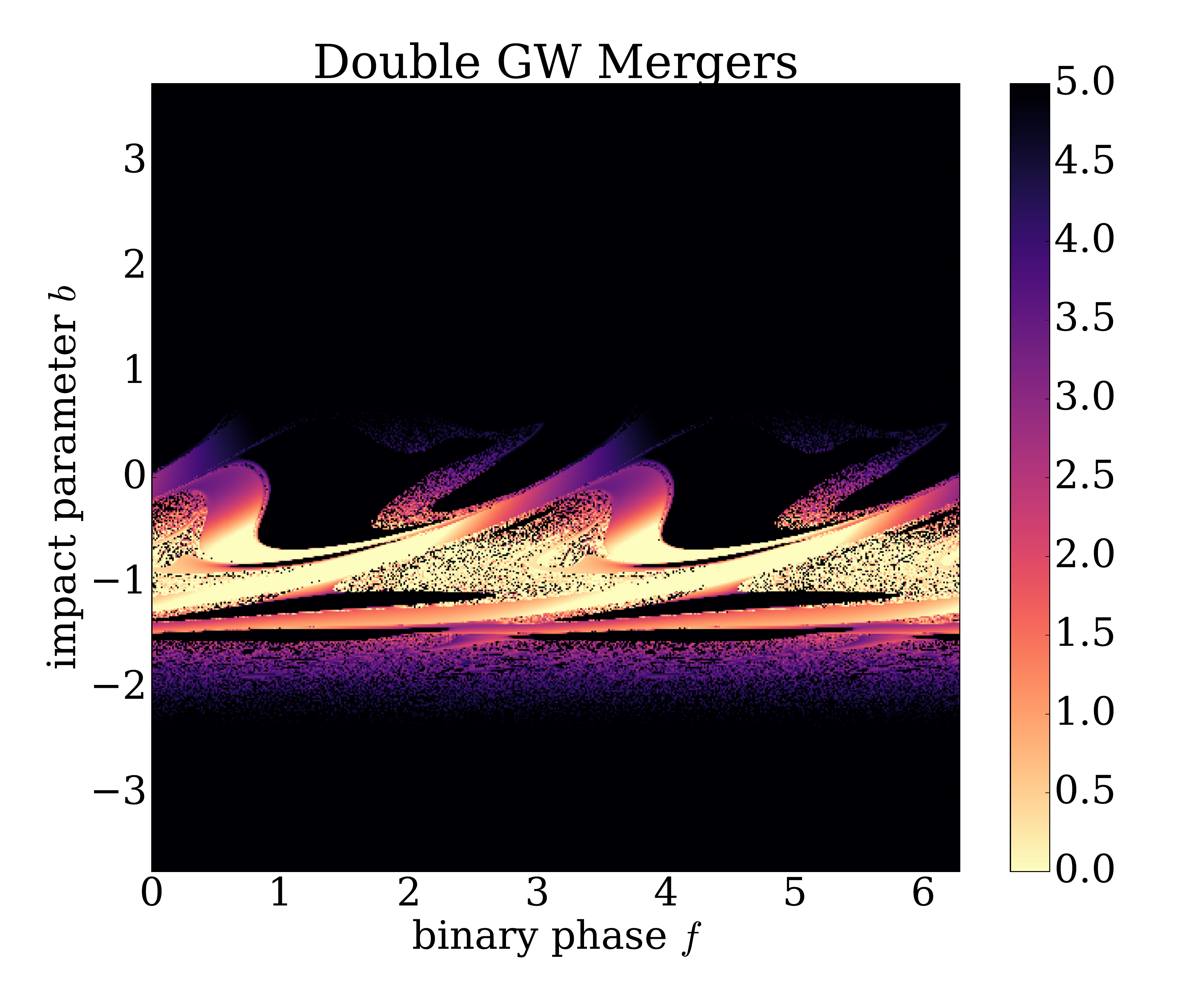}
\caption{Distribution of $\log({\Delta}t_{12}/T_{0})$, where ${\Delta}t_{12}$ is the time span from the first to the
second GW merger, and $T_{\rm 0}$ is the orbital time of the initial target binary,
derived for $a_{0} = 10^{-4}$ AU, $\gamma = 0$. As described in Section \ref{sec:Time from First to Second GW Merger},
the horizontal band for which ${\Delta}t_{12}$ takes its minimum value corresponds to where the total angular momentum of the three-body system
equals zero.
}
\label{fig:12merger}
\end{figure}

The distribution of $\log({\Delta}t_{12}/T_{0})$, where $T_{\rm 0}$ is the orbital time of the initial target binary,
as a function of $f,\ b$ for $\gamma = 0$, is shown in Figure \ref{fig:12merger}. For this we have included
all GW inspirals and BH collisions that form with a bound companion, where ${\Delta}t_{12}$ is estimated using the equations from \cite{Peters:1964bc}.
To keep this primary study simple and clear, we have not incorporated effects related to relativistic mass losses or asymmetric GW emission;
corrections that in general would lead to an increase in ${\Delta}t_{12}$.

As seen on the figure, the minimum value for ${\Delta}t_{12}$ is highly sensitive to $b$, but not $f$. The reason is that only
$b$ relates to the total angular momentum, which in turn directly relates to the corresponding GW inspiral time \citep{Peters:1964bc}.
This further explains why the value for $b$ that minimizes ${\Delta}t_{12}$ is simply the value for which the total three-body angular momentum
equals zero. As derived in Section \ref{sec:Initial Conditions}, the angular momentum of the single and the binary cancels out for
$b \approx -0.87$, which clearly is where ${\Delta}t_{12}$ takes its minimum as confirmed by the simulation results shown in the figure.
Close to this value of $b$ the three objects can, as a result of the vanishing angular momentum,
promptly merge without any significant use of GW emission. This leads to values of ${\Delta}t_{12}$ that are of order the orbital time of the initial binary,
instead of its initial GW inspiral time.

The angular momentum cancelation is less efficient in out-of-plane interactions, which explains
why the co-planar interaction case is the relevant channel to consider for generating double GW mergers.
This strongly motivates further studies with GR effects included in the $N$-body EOM of co-planar few-body interactions taking
place in AGN environments.

\section{Conclusions}\label{sec:Conclusions}

In this paper, we explored how the outcomes of binary-single BH interactions distribute
as a function of the ICs. For this we studied the corresponding graphical representation of the distribution (Figure \ref{fig:topmap_1exam}) -- 
a representation we loosely refer to as the topology in analogy with \cite{1983AJ.....88.1549H}.
Our study shows for the first time how this binary-single topology changes when GR corrections and BH finite sizes are consistently included
in the $N$-body EOM. Such corrections give rise to a non-negligible distribution of BH collisions and GW inspirals
that is directly observable by LIGO \citep[e.g.][]{2014ApJ...784...71S, 2017ApJ...840L..14S}.

Using zoom-in simulations, we resolved for the first time the emergence of micro-topological structures
residing inside RI regions (Figure \ref{fig:micro_top_struc}). By visually
inspecting orbits near these structures, we found that many of the interactions (in the RI region centered around $f \approx 1.3,\ b\approx 1.25$)
start out in a similar dynamical way characterized by a particular ordered and symmetrical orbital evolution (Figure \ref{fig:orbit_ex_1}).
To our knowledge, the classification and assembly of such symmetrical orbital evolutions is still unexplored, and a systematic
search for these throughout the $f,\ b$ space could therefore be of potential interest.

Finally, we pointed out that the inclusion of GW emission in the $N$-body EOM naturally leads to interactions resulting in two GW mergers. In this regard,
we described how the co-planar case can lead to scenarios where the time span from the first to the second GW merger is
short enough for LIGO to observe both events. This double GW merger channel could have profound astrophysical consequences,
and will therefore be explored in greater detail in an upcoming paper.

From an astrophysical perspective, our most relevant result is that the distribution of outcomes of binary-single interactions is
highly dependent on the ICs. While this certainly applies to small variations to the ICs, due to the chaotic nature of the three-body problem, we surprisingly find that
also large scale variations to the ICs give rise to corresponding large scale differences in the outcome distribution.
For example, in the co-planar case ($\gamma = 0$) the distribution is clearly different between prograde ($b>0$) and retrogade ($b<0$) interactions (Figure \ref{fig:topmap_1exam}).
This also applies to how the relative number of BH mergers distributes with the ICs (Figure \ref{fig:collGW_map}). Clear differences are also seen when
the orbital-plane angle $\gamma$ is varied (Figure \ref{fig:out_of_plane_int}).
This means that if properties of these distributions, such as the eccentricity distribution at merger,
could be observed by LIGO, then one would be able to roughly deduce the underlying distribution of ICs, which could give important hints about
the astrophysical origin. As an example, in AGN environments \citep[e.g.][]{2012MNRAS.425..460M, 2016ApJ...819L..17B, 2017arXiv170207818M}, the interactions will not only
be close to co-planer, but they will also likely be unequally distributed between prograde or retrograde interactions, depending on
the formation history of the interacting BHs and their interaction with the disc.
Our idealized study indicates that the outcome distribution is clearly different for these cases.
A natural extension to our presented work could therefore be to derive observables related to BH collisions, GW inspirals, and high eccentricity BH mergers,
for different scenarious including retrograde interactions, prograde interactions, in-plane interactions, out-of-plane interactions, and for varying mass hierarchy.

Although our paper presents details to a level that is not directly observable,
we do note that similar studies on how the distribution of BH mergers and related observables change with the ICs,
are likely to play a key role in distinguishing different astrophysical GW merger channels from each other.
An exercise that will become highly relevant when the number of observed GW mergers starts to grow.


\acknowledgments{Support for this work was provided by NASA through Einstein Postdoctoral
Fellowship grant number PF4-150127 awarded by the Chandra X-ray Center, which is operated by the
Smithsonian Astrophysical Observatory for NASA under contract NAS8-03060.}


\bibliographystyle{apj}


\end{document}